\documentclass[11pt]{article}

\usepackage{jheppub}
\usepackage[english]{babel}
\usepackage[latin1]{inputenc}
\usepackage{slashed}
\usepackage{amsmath}
\usepackage{hyperref}
\usepackage{comment}
\usepackage{physics}
\usepackage{amsthm}
\usepackage{amssymb}
\usepackage{graphicx}
\usepackage{fancyhdr}

\newcommand{\as}{\alpha_s}
\newcommand{\sss}{\scriptscriptstyle\rm}
\newcommand{\kt}{k_{\sss T}}

\newcommand{\pt}{p_{\sss T}}
\newcommand{\mur}{\mu_{\sss R}}
\newcommand{\muf}{\mu_{\sss F}}
\newcommand{\Ord}{\mathcal{O}}
\newcommand{\Cf}{C_{\sss F}}
\newcommand{\Ca}{C_{\sss A}}
\newcommand{\beq}{\begin{equation}}
\newcommand{\eeq}{\end{equation}}
\renewcommand{\(}{\left(}
\renewcommand{\)}{\right)}
\newcommand{\MSbar}{$\overline{\rm MS}$}

\definecolor{darkgreen}{RGB}{0,200,0}

\allowdisplaybreaks

\preprint{TIF-UNIMI-2017-9, MITP/17-073}

\title{High Energy Resummation of Jet Observables}

\author[a]{Simone Zoia}
\author[b,c]{and Claudio Muselli}

\affiliation[a]{PRISMA Cluster of Excellence, Johannes Gutenberg University, 55128 Mainz, Germany}
\affiliation[b]{Tif Lab, Dipartimento di Fisica, Universit\`a  degli Studi di Milano, via Celoria 16, I-20133 Milano, Italy}
\affiliation[c]{INFN, Sezione di Milano, via Celoria 16, I-20133 Milano, Italy}

\emailAdd{zoia@uni-mainz.de}
\emailAdd{claudio.muselli@mi.infn.it}

\abstract{In this paper we investigate the extension of high energy resummation at LL$x$ accuracy to jet observables. In particular, we present the high energy resummed expression of the transverse momentum distribution of the outgoing parton in the general partonic process $g(q) + g(q) \to g(q) + X$. In order to reach this result, several new ideas are introduced and exploited. First we prove that LL$x$ resummation is achieved by dressing with hard radiation an off-shell gluon initiated LO process even if its on-shell limit is vanishing or trivial. Then we present a gauge-invariant framework where these calculations can be performed by using the modern helicity techniques. Finally, we show a possible way to restore gluon indistinguishability in the final state, which is otherwise lost in the resummation procedure, at all orders in $\as$ at LL$x$. 
All partonic channels are then resummed and cross-checked against fixed-order calculations up to $\Ord\left(\as^3\right)$.}


\begin{document}

\maketitle

\flushbottom

%%%%%%%%%%%%%%%%%%%%%%%%%%%%%%%%%%%%%%%%%%%%%%%%%%%%%%%%%%%%%%%%%%%%%%%
%%%%%%%%%%%%%%%%%%%%%%%%%%%%%%%%%%%%%%%%%%%%%%%%%%%%%%%%%%%%%%%%%%%%%%%
%%%%%%%%%%%%%%%%%%%%%%%%%%%%%%%%%%%%%%%%%%%%%%%%%%%%%%%%%%%%%%%%%%%%%%%

\section{Introduction}
\label{sec:Introduction}

Among all resummation theories, high energy (or small-$x$) resummation certainly stands out by its complexity. Indeed, while other resummation techniques --~such as soft resummation~-- have been pushed up to N$^3$LL accuracy by now, high energy resummation is only available up to the first non-trivial logarithmic order. Furthermore, we have only been able to perform this kind of resummation with final states which are not strong interacting so far. 

It is now more than forty years since the first pioneering works~\cite{Lipatov:1976, Fadin:1975} made it possible to determine the singular small-$x$ contributions to the DGLAP evolution at all orders in the strong coupling $\alpha_s(Q^2)$, where $Q^2$ is the so-called hard scale of the process, and $x$ is the ratio of the latter to the partonic center-of-mass energy $\hat{s}$
\begin{align}
x = \dfrac{Q^2}{\hat{s}}.
\end{align}
It was only several years later that high energy resummation reached the first non-trivial logarithmic order accuracy~\cite{Catani:1990, CataniHautmann,CataniCiafaloniHautmann}, and anyway for inclusive cross sections alone. 

Recently, however, a new, but equivalent, approach to high energy factorization~\cite{Caola:2010} has led to the extension of this formalism to less inclusive observables: rapidity distributions first~\cite{Caola:2010}, and then transverse momentum distributions~\cite{Muselli:2015}. Besides the mere theoretical interest, these results opened up the possibility of describing final states in a more exclusive way in the context of high energy resummation, thus making this theory more appealing from the phenomenological point of view.

Moreover, the last few years have also seen significant developments in the study of coloured final states within the high energy resummation formalism. A few techniques have been proposed (e.g. see Refs.~\cite{Hameren, HamerenKotkoKutak, Kotko}), which allow calculations of processes with coloured final states to be performed in a much more efficient way than the general framework given in terms of Lipatov's effective action~\cite{Lipatov1}.
Given all such achievements, the time is ripe to tackle the extension of high energy resummation to jet observables.

This paper represents the first step towards this goal: our target will be the evaluation of the LL$x$ behaviour of the gluon or quark transverse momentum distribution in the partonic processes
\begin{align}
g(q)+g(q) \to g + X,\\
g(q)+q \to q + X,
\end{align}
where $X$ is any extra partonic radiation. Note however that, since in the high energy limit only gluon emission is not suppressed at LL$x$, the number of quarks (or anti-quarks) must be the same in both the initial and the final state. Moreover, throughout all this paper, we shall denote by $q$ the singlet quark combination, which is the only combination coupled to gluons and thus the only one exhibiting singular small-$x$ behaviour. For a discussion of the relation between singlet/non-singlet basis and standard quark/anti-quark basis we refer for instance to Refs.~\cite{Moch:2004pa,Vogt:2004mw}.

All these partonic processes have been computed at fixed order up to NNLO in Refs.~\cite{EllisSexton,Nagy:2003tz,Currie:2016bfm} in a fully exclusive calculation. Here, however, we are going to cross-check our LL$x$ predictions against fixed order results only up to NLO. At NNLO, in fact, the comparison is not yet possible, since the evaluation of Ref.~\cite{Currie:2016bfm} is only performed numerically, and no partonic distribution can be extracted in analytic form.  

It is worth noting that our parton level result could in principle be used to resum several jet observables, such as the one-jet inclusive cross section or the leading jet $\pt$ distribution. Furthermore, by using the definition of fragmentation functions, even the one-hadron inclusive cross section is suited to be investigated by means of this technique.
Here, however, we will only be concerned with the theoretical development: we will highlight and solve the technical and computational problems arising from the final state being coloured; any phenomenological analysis is left to future research. 

The interest in the generalization of high energy resummation to coloured final state is not only theoretical. Indeed, the main motivation is currently the improvement of PDFs fits. It has recently been shown that supplying fixed-order calculations with resummation effects leads to a general improvement in the theoretical accuracy of the PDF fits in the relevant kinematical region~\cite{Bonvini:2016wki, Ball:2017otu}. On top of that, historically, the gluon distribution function is known to be most constrained by data coming from jet production and DIS. Since high energy resummation for DIS is up to now well established~\cite{CataniHautmann, Bonvini:2017ogt}, high energy resummation of jet production will represent a crucial step towards small-$x$ resummed global PDFs fits.

This kind of resummation has also proven to be a very powerful tool to infer general properties of a certain process, e.g. the effect of quark masses on the Higgs transverse momentum spectrum~\cite{Caola:2016}. Moreover, small-$x$ resummed observables predict parts of the exact results, which might even be still unknown at fixed order; given the complexity of such calculations, any cross-check is most precious.

The paper is organized as follows. After a brief review of high energy resummation of transverse momentum distributions in Section~\ref{sec:HighEnergyResummation}, in Section~\ref{sec:ANewApproach} we will present a new simplified approach to the evaluation of the off-shell processes, and discuss the issue of their gauge invariance. Then, Section~\ref{sec:g*g*g} will be devoted to the evaluation of all the required off-shell cross sections. The computation will be performed using three different techniques as a matter of cross-check. Finally, the impact factors of all the parton processes will be collected in Section~\ref{sec:Results}, together with the corresponding small-$x$ resummed transverse momentum distributions, which will be cross-checked with the exact fixed-order results up to order $\alpha_s^3$. Finally, conclusions are drawn in Section~\ref{sec:Conclusions}.

%%%%%%%%%%%%%%%%%%%%%%%%%%%%%%%%%%%%%%%%%%%%%%%%%%%%%%%%%%%%%%%%%%%%%%%
%%%%%%%%%%%%%%%%%%%%%%%%%%%%%%%%%%%%%%%%%%%%%%%%%%%%%%%%%%%%%%%%%%%%%%%
%%%%%%%%%%%%%%%%%%%%%%%%%%%%%%%%%%%%%%%%%%%%%%%%%%%%%%%%%%%%%%%%%%%%%%%

\section{High Energy Resummation of Transverse Momentum Distributions}
\label{sec:HighEnergyResummation}

In this section we shall review the high energy factorization theorem for transverse momentum distributions in the leading-logarithms (LL$x$ for short) approximation. Here we will only present the most relevant aspects, basically in order to define the notation and to set the theoretical background for the following arguments. For a thorough discussion of this topic we refer the interested reader to the original paper Ref.~\cite{Muselli:2015}.

Let us consider the production of a certain state $\mathcal{S}$ with momentum $p^{\mu}$ in an hadronic collision characterized by some hard scale $Q$ and partonic center-of-mass energy $\hat{s}$. Without loss of generality, we are first going to consider the gluon-initiated process
\begin{align}
g(p_1) + g(p_2) \rightarrow \mathcal{S}(p)  + X,
\end{align}
where $X$ stands for any possible final state, and $p_1^{\mu}$ and $p_2^{\mu}$ are the momenta of the incoming gluons. In this section we will limit ourselves to the case where $\mathcal{S}$ is a colourless system. The extension to coloured final states will be discussed in the next section. 

\begin{figure}[t!]
\centering
\includegraphics[scale=0.70]{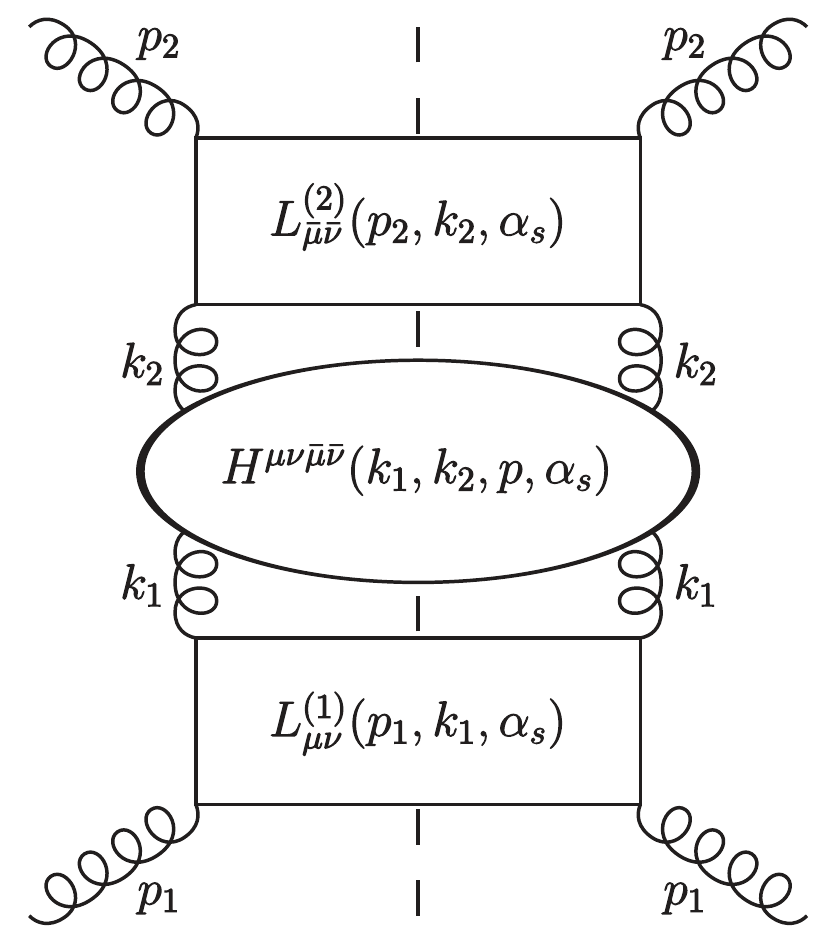}
\caption{High energy factorization of gluon-initiated hadro-production.}
\label{fig:Decomposition}
\end{figure}

The main insight of Ref.~\cite{CataniHautmann,CataniCiafaloniHautmann, EllisGeorgiMachacek,Caola:2010, Muselli:2015}, as is well known, is that, on top of collinear factorization of the hadronic observable, the partonic observable can in turn be factorized at LL$x$ accuracy. This is done by rearranging the cut diagrams contributing to the partonic observable in order to single out the so-called hard part $H^{\mu \nu \bar{\mu} \bar{\nu}}$, i.e. the 2GI kernel which is connected to the desidered final state $\mathcal{S}$ and is thus process-dependent. What is left of the cut diagrams can be cast in the form of two generally 2GR kernels $L^{(1)}_{\mu \nu}$ and $L^{(2)}_{\bar{\mu} \bar{\nu}}$, called ladder parts, which are built by iterative insertion of a suitable kernel, in the spirit of the generalized ladder expansion of Ref.~\cite{CurciFurmanskiPetronzio}. The kernel is nothing but the DGLAP anomalous dimension resummed at all orders in $\as$ at LL$x$. A practical implementation of this object can be obtained by exploiting the duality relation between BFKL and DGLAP evolution (see Refs.~\cite{AltarelliBallForte,Altarelli:1999vw,Altarelli:2001ji,Bonvini:2016wki,Bonvini:2017ogt} for further details). The ensuing diagrammatic structure is depicted in Figure~\ref{fig:Decomposition}.

The resulting factorized expression of the dimensionless transverse momentum distribution at LL$x$ accuracy is
\begin{align}
\label{eq:DecompositionpT}
Q^2 \dfrac{d\bar{\sigma}}{d \pt^2} = \int & \dfrac{Q^2}{2 \hat{s}} H^{\mu \nu \bar{\mu} \bar{\nu}} L^{(1)}_{\mu \nu} L^{(2)}_{\bar{\mu} \bar{\nu}} \delta \left( \dfrac{\pt^2}{Q^2}- \dfrac{k_{1,\sss T}^2}{Q^2}-\dfrac{k_{2,\sss T}^2}{Q^2} - 2\sqrt{\dfrac{k_{1,\sss T}^2 k_{2,\sss T}^2}{Q^2}} \cos \theta   \right) \left[ dk_1 \right] \left[ dk_2 \right],
\end{align}
where $\bar{\sigma} = Q^2 \hat{\sigma}$ is the dimensionless partonic inclusive cross section, and all dependencies have been dropped in order to simplify the notation.

A few comments on Eq.~\eqref{eq:DecompositionpT} are in order: $[d k_1]$ and $[dk_2]$ denote the relevant volume elements of the loop integrations over the gluon lines connecting the ladders to the hard part, $1/(2 \hat{s})$ is the conventional flux factor, and the apparent factors of $Q^2$ are simply meant to make the observable dimensionless. Then, $k_{1,\sss T}^{\mu}$ and $k_{2,\sss T}^{\mu}$ are the purely transverse spacelike components of $k_1^{\mu}$ and $k_2^{\mu}$, and $\theta$ is the angle between their directions. Finally, it is worth stressing that the phase space volume element of $\mathcal{S}$ is included in the hard part $H^{\mu \nu \bar{\mu} \bar{\nu}}$, whereas $L^{(1)}_{\mu \nu}$ and $L^{(2)}_{\bar{\mu}\bar{\nu}}$ contain the phase space of the ladder emissions. 

Before discussing the ensuing resummation formula, it is convenient to simplify the notation. The all-order resummation of the LL$x$ contributions is performed in Mellin space. Then, by a slight abuse of notation, we shall denote the Mellin transform of some suitable function $f(z)$ with the same symbol
\begin{align}
f\left(N \right) \equiv \mathcal{M}_{z}\left[f\right](N) = \int_0^1 d z \ z^{N-1} f\left(z\right),
\end{align}
and we shall distinguish between the two objects by their arguments.

The factorized expression Eq.~\eqref{eq:DecompositionpT} would in principle be dependent on the renormalization and factorization scales. At LL$x$ accuracy, however, the $\as$ coupling can be considered as fixed, as all the dependence on $\mur^2$ turns out to be subleading.

On the other hand, IR divergences are present in both the hard and the ladder parts, and must thus be subtracted at all orders into renormalized PDFs. Factorization of IR divergences in the hard part is straightforward, although not trivial, and will not be discussed in this paper. For details about this topic we refer the interested reader to Ref.~\cite{Marzani:2008}. Here, in fact, we shall only consider hard parts which are 2PI, rather than 2GI, and thus finite. The iterative subtraction of IR divergences in the ladders is performed using the generalized ladder expansion of Ref.~\cite{CurciFurmanskiPetronzio,Caola:2010}, and ultimately allows the resummation of the LL$x$ contributions. Once again we refer to the original references~\cite{Caola:2010, Muselli:2015}, and we present only the final resummed formula for a generic transverse momentum distribution.

Then, by introducing the dimensionless variables
\begin{align}
\label{eq:xivariables}
\xi_p = \dfrac{\pt^2}{Q^2}, \qquad \xi_1 = \dfrac{k_{1,\sss T}^2}{Q^2}, \qquad  \xi_2 = \dfrac{k_{2,\sss T}^2}{Q^2},
\end{align}
the resummed transverse momentum distribution can be expressed as
\begin{align}
\label{eq:resfact}
Q^2\frac{d\bar{\sigma}}{d\pt^2}\(N,\xi_p,\as,\frac{\muf^2}{Q^2}\)&=\gamma\(N,\as\)^2 R\left(\gamma\(N,\as\right)\)^2 e^{2\gamma\(N,\as\)\ln\frac{\muf^2}{Q^2}}\times\notag\\
&\times \int_0^\infty d\xi_1\, \xi_1^{\gamma\(N,\as\)-1}\int_0^\infty d\xi_2\, \xi_2^{\gamma\(N,\as\)-1} C_{\pt}\(N,\xi_1,\xi_2,\xi_p,\as\),
\end{align}
where we denote by $\gamma\(N,\as\)$ the LL$x$ DGLAP anomalous dimension and we define
\begin{align}
\label{eq:coeffunction}
C_{\pt}\(N,\xi_1,\xi_2, \xi_p,\as\)&\equiv \int \frac{d\theta}{2\pi} \frac{d\bar{\theta}}{2\pi} \left[\mathcal{P}^{\mu\nu} \mathcal{P}^{\bar{\mu}\bar{\nu}} H_{\mu\nu\bar{\mu}\bar{\nu}}\right]\notag\\
&\qquad\qquad\qquad\qquad\delta\(\xi_p -\xi_1-\xi_2-2\sqrt{\xi_1 \xi_2}\cos\theta\),
\end{align}
with
\begin{align}
\label{eq:proj}
\mathcal{P}^{\mu\nu}&=\frac{k_{1,\sss T}^{\mu} k_{1,\sss T}^{\nu}}{k_{1,\sss T}^{2}}, & \mathcal{P}^{\bar{\mu}\bar{\nu}}&=\frac{k_{2,\sss T}^{\bar{\mu}}k_{2,\sss T}^{\bar{\nu}}}{k_{2,\sss T}^2}.
\end{align}

The hard coefficient function $C_{\pt}$ owns a very natural physical interpretation. Indeed, it is nothing but the (dimensionless) transverse momentum distribution of the tree-level off-shell partonic process
\begin{align}
g^*(k_1) + g^*(k_2) \rightarrow \mathcal{S}(p),
\end{align}
namely the production of the desired final state $\mathcal{S}$ by fusion of two off-shell gluons. In Eq.~\eqref{eq:coeffunction} the momenta of the incoming gluons are parametrized in terms of longitudinal and transverse components as
\begin{align}
\label{eq:Kinematics}
k_i^{\mu} = z_i n_i^{\mu} + k_{i,\sss T}^{\mu}, \quad \forall i =1,2,
\end{align}
with
\begin{align}
n_i^2 = 0, \qquad \quad k_i^2 = - k_{i,\sss T}^2 <0, \qquad \quad n_i \cdot k_{j,\sss T}= 0 \quad \forall i,j=1,2.
\end{align}
The longitudinal projectors Eqs.~\eqref{eq:proj} can thus be viewed as the polarization sums for the incoming off-shell gluons, namely
\begin{align}
\label{eq:PolarizationSum}
\frac{1}{2} \sum_{\text{off-shell}} \epsilon^{\mu}(k) \epsilon^{\nu *}(k) \equiv \mathcal{P}^{\mu \nu}(k) = \dfrac{\kt^{\mu} \kt^{\nu}}{\kt^2},
\end{align}
where $\epsilon^{\mu}(k)$ denotes the polarization vector of some off-shell incoming gluon carrying momentum $k^{\mu}$ parametrized as in Eq.~\eqref{eq:Kinematics}.

Usually, in literature, Eq.~\eqref{eq:resfact} is more conveniently expressed in terms of the so-called  \emph{impact factor}
\begin{align}
\label{eq:hpTdef}
h\left(N, M_1, M_2, \xi_p, \alpha_s, \frac{\muf^2}{Q^2} \right) = \ & M_1 M_2 R(M_1) R(M_2) e^{\(M_1+M_2\)\ln\frac{\muf^2}{Q^2}} \times \nonumber \\
& \times \int_0^{\infty} d\xi_1 \xi_1^{M_1-1} \int_0^{\infty} d\xi_2 \xi_2^{M_2-1} C_{\pt}(N, \xi_1, \xi_2, \xi_p, \alpha_s),
\end{align}
so that 
\beq
\label{eq:relres}
Q^2\frac{d\bar{\sigma}}{d\pt^2}\(N,\xi_p,\as,\frac{\muf^2}{Q^2}\)= h_{\pt}\(N,\gamma\(N,\as\), \gamma\(N,\as\),\xi_p,\as,\frac{\muf^2}{Q^2}\).
\eeq

The factor $R(M)$ in Eqs.~\eqref{eq:resfact} and~\eqref{eq:hpTdef} is a function which contains the entire dependence on the factorization scheme. In particular, $R(M)$ was calculated in the \MSbar~scheme in Ref.~\cite{CataniHautmann}, and reads 
\begin{align}
\label{eq:R(M)}
R(M) & = \left\{\dfrac{\Gamma(1-M) \chi_0(M)}{\Gamma(1+M) \left[-M \chi_0'(M) \right]} \right\}^{\frac{1}{2}} \exp\left\{M \psi(1)+\dfrac{\Ca}{\pi} \int_0^M dM' \dfrac{\psi'(1)-\psi'(1-M')}{\chi_0(M')} \right\} = \nonumber \\
& = 1 + \dfrac{8}{3} \zeta(3) M^3+ \mathcal{O}\left(M^4\right),
\end{align}
where $\Ca = 3$, $\chi_0(M)$ is the LO BFKL kernel
\begin{align}
\label{eq:chi0}
\chi_0(M) = \dfrac{\Ca}{\pi}  \left(2 \psi(1)-\psi(M)-\psi(1-M)\right),
\end{align}
$\psi(M)$ the Digamma function, and $\zeta(n)$ the Riemann $\zeta$-function.

Up to now, we have only considered gluon initiated channels. However, due to the fact that the quark entries of the full resummed anomalous dimension matrix only give NLL$x$ contributions~\cite{AltarelliBallForte}, the high energy resummed distributions of the other partonic channels of production of $\mathcal{S}$ can be derived at LL$x$ accuracy from that of the purely gluonic channel through the trivial relations
\begin{align}
\label{eq:Relationqg}
\dfrac{d\bar{\sigma}_{\text{res}}}{d\xi_p}\biggl|_{qg} & = \dfrac{\Cf}{\Ca} \left[h\left(N, \gamma_s\left(\frac{\alpha_s}{N}\right), \gamma_s\left(\frac{\alpha_s}{N}\right), \xi_p, \alpha_s \right)- h\left(N, \gamma_s\left(\frac{\alpha_s}{N}\right), 0, \xi_p, \alpha_s \right) \right], \\
\label{eq:Relationqq}
\dfrac{d\bar{\sigma}_{\text{res}}}{d\xi_p}\biggl|_{qq} & = \left(\dfrac{\Cf}{\Ca}\right)^2 \left[h\left(N, \gamma_s\left(\frac{\alpha_s}{N}\right), \gamma_s\left(\frac{\alpha_s}{N}\right), \xi_p, \alpha_s \right)- 2 h\left(N, \gamma_s\left(\frac{\alpha_s}{N}\right), 0, \xi_p, \alpha_s \right) \right],
\end{align}
with $\Ca = 3$ and $\Cf = 4/3$.\footnote{See Ref.~\cite{Muselli:2015} for details.}

%%%%%%%%%%%%%%%%%%%%%%%%%%%%%%%%%%%%%%%%%%%%%%%%%%%%%%%%%%%%%%%%%%%%%%%
%%%%%%%%%%%%%%%%%%%%%%%%%%%%%%%%%%%%%%%%%%%%%%%%%%%%%%%%%%%%%%%%%%%%%%%
%%%%%%%%%%%%%%%%%%%%%%%%%%%%%%%%%%%%%%%%%%%%%%%%%%%%%%%%%%%%%%%%%%%%%%%

\section{A New Prospective in Off-shell Calculations}
\label{sec:ANewApproach}

In the previous section we have seen that, in the case of hadro-production, the partonic observables we are interested in --~namely inclusive cross sections and transverse momentum distributions~-- can be decomposed into a hard part and two ladders at LL$x$ accuracy. The hard part is formed by the LO observable evaluated with the two initial gluons put off-shell, whereas the ladder parts are computed by iteration of a proper emission kernel. This picture was derived for colourless final states, but no change is necessary if the final state is instead coloured. However, calculations become more difficult from the technical point of view, and some caution is needed.

First we have to properly define the off-shell quantity required in order to perform high energy resummation with the desidered coloured final state. Then, we need to present an efficient procedure to actually compute it.

For simplicity, we shall for now limit ourselves to the pure gluonic channel. Therefore, we want to evaluate at all orders in $\as$ the high energy behaviour of the process
\beq
g + g \to  g + X,
\eeq
where $X$ is a bunch of extra emitted gluons. Since the first non-trivial contribution to this process is
\beq
\label{eq:LOgg}
g + g \to g + g,
\eeq
one could naively think that its off-shell version, namely
\beq
g^* + g^* \to g + g,
\eeq
would be the best candidate for the evaluation of the hard part.

However, this turns out not to be the case. This picture, which is a legacy of the first applications of high energy resummation to inclusive cross sections, is in fact in contrast with the theoretical derivation based on the generalized ladder expansion of Ref.~\cite{Caola:2010, Muselli:2015}, sketched out in the previous section. 

Following the general treatment of Section~\ref{sec:HighEnergyResummation}, in fact, the hard coefficient function $C_{\pt}$ has to be computed by considering the tree level process which produces the desired final state $\mathcal{S}$ -- either colourless or coloured -- by fusion of two initial off-shell gluons. In our case, then, it is clear that the relevant off-shell process should be
\beq
\label{eq:LOg}
g^* + g^* \to g,
\eeq
rather than Eq.~\eqref{eq:LOgg}.

Even if, at first sight, this principle could seem not very straightforward, it has already been applied in some sense in the high energy differential calculation of Refs.~\cite{Caola:2010, Muselli:2015} for the Higgs boson production. In Ref.~\cite{Muselli:2015}, for example, the authors use the off-shell process
\begin{align}
g^* + g^* \rightarrow H
\end{align}
to evaluate the hard coefficient function, even though the corresponding transverse momentum distribution is trivial in the on-shell limit. 
Indeed, in order for the Higgs boson to have a non-vanishing transverse momentum, the final state has to include at least a second recoiling particle, which at LL$x$ accuracy can only be a gluon. Thus, the series expansion of the resummed result in powers of $\alpha_s$ vanishes at $\mathcal{O}(\alpha_s^2)$, and the first non-trivial prediction only appears at $\mathcal{O}(\alpha_s^3)$, with the LL$x$ contribution to the transverse momentum distribution of
\begin{align}
g + g \rightarrow H + g.
\end{align}

Therefore, we shall present a new criterion for high energy calculations, based on new analysis carried out in the context of high energy resummation of transverse momentum distributions. The hard part has to be identified with the \emph{LO off-shell observable}, even if the corresponding on-shell limit is vanishing or trivial. In such cases, the resummation procedure will guarantee that at least one extra emission from the ladders is present. This criterion will substantially simplify our calculations, enabling us to reach a very compact final result.

A last remark is however necessary. Note that, in general, the inclusion of extra radiation in the hard part is not a priori incorrect, but only highly disadvantageous. In fact, it would lead to a new resummed prediction, different only for subleading contributions, but at the cost of raising considerably the difficulty of the actual computation. 

This point is closely linked to the second issue we need to tackle when we want to evaluate off-shell quantities with coloured final states: the intrinsic computational difficulty of these calculations.   

It is indeed well known that, in certain cases, computation by means of standard Feynman diagrams might be unnecessarily cumbersome, since the cancellation between physical and unphysical gluon polarizations only occur at the level of the squared modulus, and not directly in the amplitude.
For this reason, in recent years, modern and more efficient techniques --~such as the helicity formalism reviewed e.g. in Refs.~\cite{PeskinColour,Dixon}~-- have been proposed, in order to take into account in the actual computation only the physical degrees of freedom. 

Such techniques basically rely on the gauge invariance of the observable which has to be computed. However, while for on-shell observables gauge invariance is immediately verified, the same is no longer true in general when working with off-shell quantities, such as the hard part in the high-energy resummation procedure. This issue is of particular importance, and certainly deserves a separate subsection.

%%%%%%%%%%%%%%%%%%%%%%%%%%%%%%%%%%%%%%%%%%%%%%%%%%%%%%%%%%%%%%%%%%%

\subsection{Restoring Gauge Invariance in the Hard Part}
\label{sec:RestoringGaugeInvariance}

First of all, it is important to stress that, from now on, what we actually mean by saying that some $n$-gluon amplitude $\mathcal{M}(\epsilon_1, ..., \epsilon_n)$, dependent on the polarization vectors $\epsilon_i$ with $i =1,...,n$, is gauge-invariant is that it satisfies the Abelian Ward identities, namely
\begin{align}
\label{eq:AbelianWardIdentity}
\mathcal{M}\left(\epsilon_1, ..., \epsilon_{i-1}, k_i, \epsilon_{i+1},..., \epsilon_n \right) = 0 \quad \forall i = 1, ..., n.
\end{align}
Indeed, these are the relations we need to be true in order to be able to exploit the modern and more efficient techniques relying on gauge invariance.

It is well known that Eq.~\eqref{eq:AbelianWardIdentity} holds in a non-Abelian gauge theory such as QCD when all but at most one of the external gluons are on-shell. It is therefore clear why this might represent an issue in the context of high energy resummation: when working with off-shell external gluons as required by the resummation procedure, gauge invariance in the sense of Eq.~\eqref{eq:AbelianWardIdentity} might be lost, and we might thus be forced to use the inefficient standard evaluation technique. If that were the case, complicated processes such as those involved in multi-jets observables might be beyond the reach of high energy resummation.

Our issue is essentially related to the computation of the hard part. Since it involves four external gluons (see Fig.~\ref{fig:Decomposition}), it turns out that the underlying off-shell amplitude does not fulfil Eq.~\eqref{eq:AbelianWardIdentity} in general. However, we can use the general studies on gauge invariance of off-shell amplitudes of Refs.~\cite{Hameren, Kotko, HamerenKotkoKutak} to restore Eq.~\eqref{eq:AbelianWardIdentity} in the case of our hard coefficient function $C_{\pt}$. 

We are thus going to change our definition of the off-shell scattering amplitude by adding gauge-dependent terms so that it fulfils Eq.~\eqref{eq:AbelianWardIdentity}. It is worth stressing that, obviously, all these extra terms will cancel each other out in any physical observable, thanks to gauge symmetry. 

In the following we will use two general set of prescriptions, presented in Refs.~\cite{Hameren} and~\cite{Kotko} respectively. We shall see that the off-shell process has to be evaluated by considering the off-shell gluons as radiated off eikonal quarks in the former technique, and straight infinite Wilson lines in the latter. In particular, we find that this last approach is very suitable to the extension to more complicated multi-jet processes, and we are thus going to briefly review its most relevant features in Appendix~\ref{sec:WilsonLines}.

The process described by Eq.~\eqref{eq:LOg} is indeed rather simple, and even computation by means of Feynman diagrams is in principle affordable. However, limiting ourselves only to the standard technique would result in restricting the scope of high energy resummation just to very simple processes. Instead, we are going to exploit the relative simplicity of the process considered in this paper as a cross-check for the more complicated gauge invariant construction of the hard part. Indeed, since no observable with coloured final state has been resummed in the high energy limit as yet, having different ways of carrying out the calculations to cross-check the results is fundamental, in order to provide a strong ground for the theory.

This we will do in the next section, where the calculation of the cross section of the off-shell process $g^*+ g^* \to g$ will be presented using three different techniques: the standard QCD Feynman rules in the Feynman Gauge,\footnote{We are forced to select the Feynman Gauge, because it is the particular gauge choice we use to perform the IR subtraction of the ladders.} and the two gauge-restoring procedures proposed in Refs.~\cite{Kotko} and~\cite{Hameren} respectively.
 
We also recall that, in order to evaluate the LL$x$ transverse momentum dependence of the outgoing quark in processes of the form $g(q) + q \to q + X$, we also need to compute the cross section of the off-shell process $g^* + q \to q$. This calculation will be postponed until Section~\ref{sec:g*qq}.

%%%%%%%%%%%%%%%%%%%%%%%%%%%%%%%%%%%%%%%%%%%%%%%%%%%%%%%%%%%%%%%%%%%%%%%
\section{Calculation of the Hard Off-shell Cross Sections}
\label{sec:g*g*g}

The basic object we need to evaluate in order to resum all gluon production channels $g(q) + g(q) \to  g + X$ in the small-$x$ limit is the scattering amplitude of the off-shell process
\begin{align}
\label{eq:g*g*g}
g^*_{c_1}(k_1) + g^*_{c_2}(k_2) \rightarrow g_c(p),
\end{align}
where the subscripts denote the colour of each gluon, and the kinematics is parametrized as in Eq.~\eqref{eq:Kinematics}. As explained in Section~\ref{sec:RestoringGaugeInvariance}, this result can be achieved in a number of ways, depending on whether --~and how~-- we want gauge invariance to be restored. 

We once again stress that there is no practical need for gauge invariance to be restored, because the calculation can be carried out with the usual QCD Feynman rules, as we shall do in Section~\ref{sec:g*g*g_FR}. Although straightforward, we shall see that even this standard calculation features some non-trivial simplifications, a priori not expected, which lead to a very compact result. 
Then, in order to provide a solid cross-check, we will repeat the same calculation first with the Wilson lines technique of Ref.~\cite{Kotko} in Section~\ref{sec:g*g*g_WL}, and next with the eikonal quark prescriptions of Ref.~\cite{Hameren} in Section~\ref{sec:g*g*g_EQ}.
  
It will be apparent that these methods, although different in both principle and practice, all lead to the same result, thus substantiating not only its correctness, but also the explicit consistency of the gauge-restoring techniques of Refs.~\cite{Kotko, Hameren}.

Once the validity of our method has been attested in the pure gluonic channel, we can move on to consider more swiftly the last ingredient needed in the resummation recipe, namely the cross section of the quark-production off-shell process $g^* q \rightarrow q$. This we shall present in Section~\ref{sec:g*qq}, were the calculation will be carried out only with the standard QCD Feynman rules. Since there is now only one external off-shell gluon leg, the consistency with the two gauge-restoring techniques is immediate, and thus shall not be discussed.

%%%%%%%%%%%%%%%%%%%%%%%%%%%%%%%%%%%%%%%%%%%%%%%%%%%%%%%%%%%%%%%%%%%%%%%

\subsection{$g^* g^* \rightarrow g$ calculation using QCD Feynman rules}
\label{sec:g*g*g_FR}

Using the standard QCD Feynman rules only, it is trivial to see that the process described by Eq.~\eqref{eq:g*g*g} receives contributions from just one diagram at tree level, namely the three-gluon vertex. The corresponding amplitude then simply reads 
\begin{align}
\label{eq:MF}
\mathcal{M}_{\text{\tiny{F}}} = & -g_s f^{\Ca c_B c} V_{\mu \nu \gamma} (k_1, k_1, -p) e^{\mu}_1 e^{\nu}_2 \epsilon^{* \gamma},
\end{align}
where $e_1^{\mu}$ and $e_2^{\mu}$ are the polarization vectors of the incoming off-shell gluons, $\epsilon^{\mu}$ is the polarization vector of the outgoing on-shell gluon, and $V_{\alpha \beta \gamma}(k_1, k_2, k_3)$ is the tensor structure of the three-gluon coupling with all momenta entering the vertex, namely
\begin{align}
\label{eq:V}
V_{\alpha \beta \gamma}(k_1, k_2, k_3) = g_{\alpha \beta}(k_1-k_2)_{\gamma} + g_{\beta \gamma}(k_2-k_3)_{\alpha} + g_{\gamma \alpha}(k_3-k_1)_{\beta}.
\end{align}
Note that the incoming off-shell momenta $k_1^{\mu}$ and $k_2^{\mu}$ have to be written in the usual high energy parametrization, given by Eq.~\eqref{eq:Kinematics}.

The calculation of the squared matrix element is then rather lengthy, but completely straightforward. One only has to make sure that the sum over the polarizations of the off-shell gluons is carried out according to
Eq.~\eqref{eq:PolarizationSum}. The sum over the polarizations of the outgoing on-shell gluon, on the other hand, is performed as usual.

Squaring Eq.~\eqref{eq:MF}, averaging over colour and polarization of the incoming particles, and summing over those of the outgoing one thus gives
\begin{align}
\label{eq:MF2}
\overline{| \mathcal{M}_{\text{\tiny{F}}}|^2} =  - g_s^2 \ \dfrac{\Ca}{8} \ \dfrac{\left|\vec{k}_{1,\sss T} + \vec{k}_{2,\sss T}\right|^2 \left( (\vec{k}_{1,\sss T} \cdot \vec{k}_{2,\sss T})^2 - k_{1,\sss T}^2 k_{2,\sss T}^2  \right) - (\vec{k}_{1,\sss T} \cdot \vec{k}_{2,\sss T})^2 z_1 z_2 \hat{s}}{k_{1,\sss T}^2 k_{2,\sss T}^2},
\end{align}
where $\hat{s}$ is the partonic center-of-mass energy
\begin{align}
\hat{s} = (n_1+n_2)^2 = 2 n_1 \cdot n_2.
\end{align}

Indeed, Eq.~\eqref{eq:MF2} is a complicated expression for a really simple result. In order to make the simplification apparent, it is first convenient to introduce the angle $\theta$ between the directions of the spatial vectors $\vec{k}_{1,\sss T}$ and $\vec{k}_{2,\sss T}$, so that
\begin{align}
\label{eq:kAkB}
( \vec{k}_{1,\sss T} \cdot \vec{k}_{2,\sss T} ) = k_{1,\sss T} k_{2,\sss T} \cos \theta.
\end{align}
Substituting Eq.~\eqref{eq:kAkB} into Eq.~\eqref{eq:MF2} then gives
\begin{align}
\label{eq:SquareAmplitudeF_tau}
\overline{\left| \mathcal{M}_{\text{\tiny{F}}}\right|^2} =  g_s^2 \ \dfrac{\Ca}{8}   z_1 z_2 \hat{s} \left(\tau \sin^2 \theta + \cos^2 \theta \right),
\end{align}	
where we have introduced the useful dimensionless variable $\tau$
\begin{align}
\label{eq:tau}
\tau = \dfrac{\pt^2}{z_1 z_2 \hat{s}},
\end{align}
which will be used throughout the rest of this paper. Moreover, momentum conservation sets $\tau = 1$, so that we are simply left with
\begin{align}
\label{eq:SquaredAmplitudeF}
\overline{\left| \mathcal{M}_{\text{F}}\right|^2} =  g_s^2 \ \dfrac{\Ca}{8}   z_1 z_2 \hat{s},
\end{align}	
which is our final expression.

Putting together the squared matrix element Eq.~\eqref{eq:SquaredAmplitudeF}, the phase space volume element
\begin{align}
\label{eq:dPhi1}
d \Phi_1 = \dfrac{2 \pi}{|\vec{p}_T|^2} \delta \left( \dfrac{1}{\tau}-1 \right),
\end{align}
and the flux factor
\begin{align}
\label{eq:phi}
\phi = 2 z_1 z_2 \hat{s}
\end{align}
finally gives the cross section of the off-shell process $g^* g^*\rightarrow g$
\begin{align}
\label{eq:sigmag*g*g}
\sigma \big|_{g^* g^* \rightarrow g} = \dfrac{\sigma_0}{|\vec{p}_T|^2} \ \delta(1-\tau),
\end{align}
where we have defined
\begin{align}
\label{eq:sigma0}
\sigma_0 = \dfrac{\Ca \pi^2}{2} \alpha_s
\end{align}
in order to simplify the notation. 

As we shall see in the next sections, although the corresponding scattering amplitudes will differ from Eq.~\eqref{eq:MF}, both the gauge-restoring techniques will lead to this same cross section, thus providing the desidered cross-check.

%%%%%%%%%%%%%%%%%%%%%%%%%%%%%%%%%%%%%%%%%%%%%%%%%%%%%%%%%%%%%%%%%%%%%%%

\subsection{$g^* g^* \rightarrow g$ calculation using Wilson lines}
\label{sec:g*g*g_WL}

\begin{figure}[t!]
\centering
\includegraphics[scale=0.85]{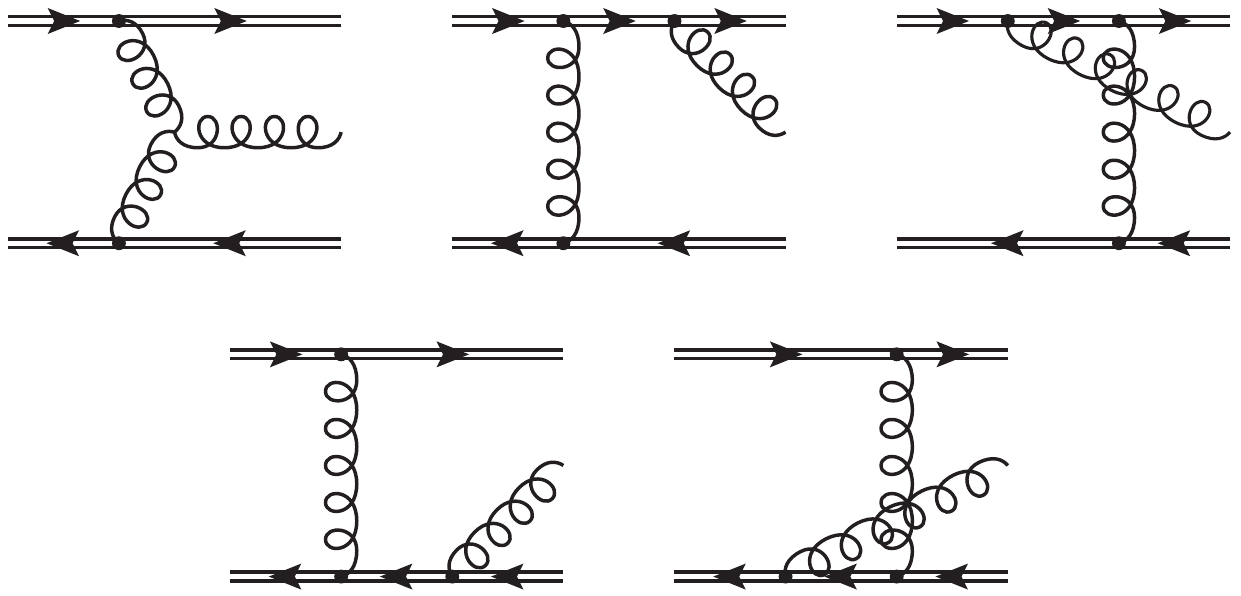}
\caption{Feynman diagrams contributing to the gauge-invariant tree-level scattering amplitude of $g^*+g^*\rightarrow g$ in the Wilson lines formalism (see Appendix~\ref{sec:WilsonLines}). The first diagram represents the standard QCD contribution; the other four have to be added in order to restore gauge invariance.}
\label{fig:Mtot}
\end{figure}

\begin{figure}[t!]
\centering
\includegraphics[scale=0.65]{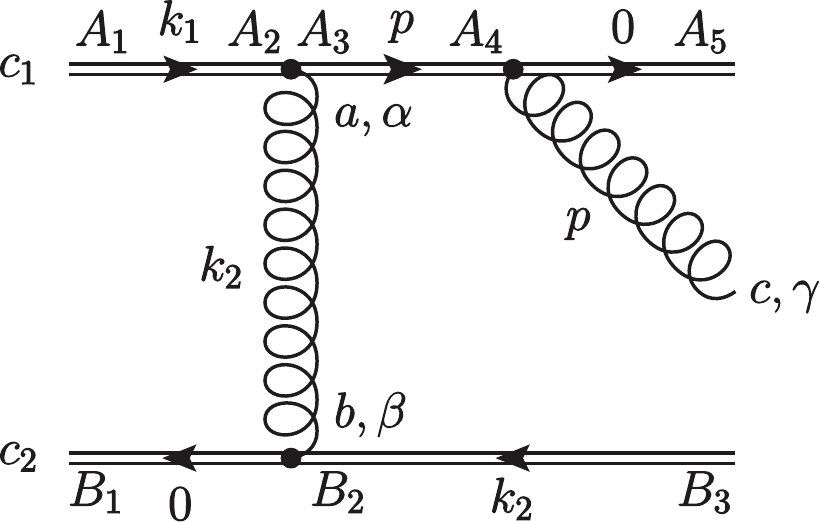}
\caption{One of the non-QCD Feynman diagrams which restore gauge invariance in the scattering amplitude of $g^*_{c_1}(k_1) + g^*_{c_2}(k_2) \rightarrow g_c(p)$ in the Wilson lines formalism. The outgoing gluon is radiated directly off one of the straight infinite Wilson lines associated with the incoming off-shell gluons.}
\label{fig:M2}
\end{figure}

With the addition of the Wilson lines Feynman rules, which we have summarised in Figure~\ref{fig:Wilson_FR} for the convenience of the reader, the gauge-invariant scattering amplitude of the off-shell process~\eqref{eq:g*g*g} receives contributions from the five diagrams depicted in Figure~\ref{fig:Mtot}. 
It is worth noting that, in this formalism, the outgoing gluon can be radiated directly off the Wilson lines representing the off-shell incoming gluons. Indeed, gauge invariance is restored from exactly this kind of diagrams, whose contributions can all be inferred, by exploiting the apparent symmetries, from that of the prototype diagram shown in Figure~\ref{fig:M2}, given by
\begin{align}
\mathcal{M}_2 = - 2 i g_s \dfrac{(e_1 \cdot e_2) (e_1 \cdot \epsilon^*)}{k_2^2 \ (p \cdot e_1)} \Tr\Bigl[T^{c_1} T^{c_2} T^c \Bigr],
\end{align}
where we have retained the same notation of the previous section for the polarization vectors, and we denote by $T^a$ the generators of the $SU(3)$ colour group in the fundamental representation, normalized as $\Tr\left[T^a T^b\right] = \delta^{ab}/2 $. 
 
Adding up all five contributions depicted in Figure~\ref{fig:Mtot} gives the full scattering amplitude, which is most conveniently written as
\begin{align}
\label{eq:MWL}
\mathcal{M}_{\text{\tiny{W}}}(\epsilon) = g_s f^{c_1 c_2 c} \mathcal{M}_{\gamma} \epsilon^{*\gamma},
\end{align}
with
\begin{align}
\label{eq:Mgamma}
\mathcal{M}_{\gamma} = - \dfrac{e_1^{\mu} e_2^{\nu} }{k_1^2 k_2^2} V_{\mu \nu \gamma}(k_1, k_2,-p) + \dfrac{(e_1 \cdot e_2) }{k_2^2 \ (p \cdot e_1)} e_{1 \gamma} - \dfrac{(e_1 \cdot e_2)}{k_1^2 \ (p \cdot e_2)} e_{2 \gamma},
\end{align}
where $V_{\alpha \beta \gamma}(k_1, k_2, k_3)$ is given by Eq.~\eqref{eq:V}.

Equation~\eqref{eq:MWL} is clearly different from the scattering amplitude obtained by means of the standard QCD Feynman rules Eq.~\eqref{eq:MF}, but it is straightforward to check that it indeed fulfils Eq.~\eqref{eq:AbelianWardIdentity}, namely that
\begin{align}
\mathcal{M}_{\text{\tiny{W}}}(p) = 0.
\end{align}

Note that, as explained in Appendix~\ref{sec:WilsonLines}, the polarization vectors of the off-shell gluons $e_i^{\mu}$ have to treated as eikonal couplings in the Wilson lines formalism. Their explicit expression, given by
\begin{align}
\label{eq:EikonalCoupling}
e_i^{\mu} = z_i k_{i,\sss T} n_i^{\mu}, \quad \forall i=1,2,
\end{align}
automatically implements the average rule Eq.~\eqref{eq:PolarizationSum}, without the need of any additional averaging factor.

Thus, we just need to average over colour of the incoming gluons, and to sum over colour and polarization of the outgoing gluon, obtaining
\begin{align}
\label{eq:M2g*g*g1}
\overline{\left|\mathcal{M}_{\text{\tiny{W}}}\right|^2} = 8 \pi \alpha_s \dfrac{\Ca}{8} \dfrac{1}{k_1^2 k_2^2} \dfrac{(e_1 \cdot e_2)^3}{(k_1 \cdot e_2) (k_2 \cdot e_1)}.
\end{align}
Substituting Eq.~\eqref{eq:EikonalCoupling} for $e_1^{\mu}$ and $e_2^{\mu}$ into Eq.~\eqref{eq:M2g*g*g1} then gives the final expression of the unpolarized and colour-averaged squared matrix element of the off-shell process $g^* g^*\rightarrow g$
\begin{align}
\label{eq:M2g*g*g}
\overline{\left|\mathcal{M}_{\text{\tiny{W}}}\right|^2} =  \alpha_s \dfrac{\Ca \pi}{2} z_1 z_2 \hat{s} \equiv \overline{\left|\mathcal{M}_{\text{\tiny{F}}}\right|^2}.
\end{align}

It is now apparent, if we compare Eq.~\eqref{eq:M2g*g*g} to Eq.~\eqref{eq:SquaredAmplitudeF}, that the standard QCD Feynman rules in the Feynman Gauge and the Wilson lines formalism both lead to the same result at the level of the squared matrix element.  
As anticipated, thus, the additional terms allowing for the scattering amplitude to fulfil the Abelian Ward identity cancel each other out when considering a physical observable.
In the next section we shall see that, as expected, this holds also for the technique of Ref.~\cite{Hameren}.

%%%%%%%%%%%%%%%%%%%%%%%%%%%%%%%%%%%%%%%%%%%%%%%%%%%%%%%%%%%%%%%%%%%%%%%%

\subsection{$g^* g^* \rightarrow g$ calculation using eikonal quark lines}
\label{sec:g*g*g_EQ}

The basic idea at the heart of the procedure presented in Ref.~\cite{Hameren} is that the off-shell process given by Eq.~\eqref{eq:g*g*g} has to be embedded into a quark-scattering process $q_i + q_m \rightarrow q_j + q_n + g_c$. The off-shell gluons are thus to be thought of as radiated off eikonal quarks, i.e. quarks to which they are attached via the eikonal coupling given by Eq.~\eqref{eq:EikonalCoupling}. As a result, the relevant diagrams have the same topologies as those depicted in Figure~\ref{fig:Mtot}, the only difference being that the Wilson lines have to be traded for eikonal quarks.

Several other adjustments need to be made in order to write down the analytic expression of the diagrams, but a full discussion of this method is beyond our scope, and we thus refer to the original paper Ref.~\cite{Hameren}, where a thorough list of the prescriptions to be followed is given. Here, instead, we just show the ensuing scattering amplitude
\begin{align}
\label{eq:totalamplitudeH}
\mathcal{M}_{\text{\tiny{Q}}}(\epsilon) = - i F g_s^3 f^{c_1 c_2 c} (T^{c_1})_{ij} (T^{c_2})_{mn}  \mathcal{M}^{'}_{\gamma} \epsilon^{*\gamma} ,
\end{align}
where we have defined
\begin{align}
\label{eq:MgammaHameren}
\mathcal{M}^{'}_{\gamma} =  - \dfrac{n_1^{\alpha} n_2^{\beta} V_{\alpha \beta \gamma} (k_1, k_2, -p)}{k_1^2 k_2^2} + \dfrac{(n_1 \cdot n_2) \ n_{1 \gamma}}{(n_1 \cdot k_2) \ k_2^2}  - \dfrac{(n_1 \cdot n_2) \ n_{2 \gamma}}{(n_2 \cdot k_1) \ k_1^2},
\end{align}
and the factor $F$ is given by Ref.~\cite{Hameren}, and reads
\begin{align}
\label{eq:F}
F = \frac{i z_1 \sqrt{2 k_{1,\sss T}^2}}{g_s} \frac{i z_2 \sqrt{2 k_{2,\sss T}^2}}{g_s}.
\end{align}

Comparing Eqs.~\eqref{eq:totalamplitudeH} and~\eqref{eq:MgammaHameren} to Eqs.~\eqref{eq:MWL} and~\eqref{eq:Mgamma}, it is no surprise that, unlike the case of the standard QCD Feynman rules, there is a strong resemblance between the two gauge-restoring techniques already at the level of the scattering amplitudes. Indeed, we note that the tensor structure is precisely the same, since 
\begin{align}
\label{eq:MgammaSimilarity}
\mathcal{M}_{\gamma} = z_1 z_2 k_{1,\sss T} k_{2,\sss T} \mathcal{M}^{'}_{\gamma}.
\end{align}
Then, substituting Eq.~\eqref{eq:MgammaSimilarity} and the explicit expression of the factor $F$ into Eq.~\eqref{eq:totalamplitudeH} gives
\begin{align}
\mathcal{M}_{\text{\tiny{Q}}}(\epsilon) = g_s \left[2 i f^{c_1 c_2 c}  \left(T^{c_1}\right)_{ij} (T^{c_2})_{mn} \right] \mathcal{M}_{\gamma} \epsilon^{*\gamma},
\end{align}
which only differs from Eq.~\eqref{eq:MWL} in the colour factor.

A bit of colour algebra then shows that full equality emerges after squaring
\begin{align}
\overline{\left|\mathcal{M}_{\text{\tiny{Q}}} \right|^2} =g_s^2 \dfrac{\Ca}{8}  \left(- \mathcal{M}^*_{\gamma} \mathcal{M}^{\gamma} \right) \equiv \overline{\left|\mathcal{M}_{\text{\tiny{W}}} \right|^2} \equiv \overline{\left|\mathcal{M}_{\text{\tiny{F}}} \right|^2}.
\end{align}
Note that, while the averaging factors associated with the incoming eikonal quarks are already included in $F$ \cite{Hameren}, the overall factor of $1/(N_c^2-1)^2$ corresponding to the average over the colours of the incoming gluons has to be put by hand.

We can thus conclude that the three techniques employed in this paper all lead to the same expression for the cross section of the off-shell process $g^*+g^*\to g$, given by Eq.~\eqref{eq:sigmag*g*g}. This solid result will then allow, in Section~\ref{sec:Results}, the resummation of all partonic channels $g(q)+g(q)\rightarrow g + X$ in the small-$x$ limit. Before moving on to that, however, let us briefly consider quark production.

%%%%%%%%%%%%%%%%%%%%%%%%%%%%%%%%%%%%%%%%%%%%%%%%%%%%%%%%%%%%%%%%%%%%%%%

\subsection{$g^* q \rightarrow q$ calculation}
\label{sec:g*qq}
The other off-shell process we need to consider is quark production. As gluon fusion can not give rise to just a quark in the final state, one of the incoming partons has to be a quark. In this section, thus, we shall discuss the calculation of the unpolarized squared matrix element of the off-shell process
\begin{align}
\label{eq:qg*q}
q_i(k_1) + g^*_a(k_2) \rightarrow q_j(p),
\end{align}
where, again, the subscripts denote the parton colours. 

The momentum of the off-shell gluon $k_2^{\mu}$ has the usual high energy parametrization given by Eq.~\eqref{eq:Kinematics}. On the other hand, since the incoming quark is on-shell, its momentum $k_1^{\mu}$ has no transverse component and is thus parametrized as 
\begin{align}
k_1^{\mu} = z_1 n_1^{\mu}, \qquad n_1^2 = 0.
\end{align}

Since there is no gluon in the final state, the Abelian Ward identities are in this case automatically fulfilled. Nonetheless, one could still carry out the calculation by means of the gauge-restoring techniques, and indeed the cross-check turns out to be successful in this case, too. Since there is now only one external off-shell gluon, however, such computations are far less interesting then those presented in Sections~\ref{sec:g*g*g_WL} and~\ref{sec:g*g*g_EQ}. Here, therefore, we shall only briefly review the calculation using the standard QCD Feynman rules in the Feynman Gauge.

The tree-level scattering amplitude then receives contributions just from the quark-gluon vertex, and simply reads
\begin{align}
\label{eq:Mqg*q}
\mathcal{M} = - g_s \left( T^a \right)_{ij} \bar{u}_j(p) \gamma_{\alpha} u_i(k_1) e^{\alpha}(k_2),
\end{align}
where $e^{\alpha}(k_2)$ is the polarization vector of the off-shell incoming gluon. Squaring Eq.~\eqref{eq:Mqg*q}, averaging over colour and polarization of the incoming particles, and summing over those of the outgoing one, gives
\begin{align}
\label{eq:M2qg*q}
\overline{\left| \mathcal{M} \right|^2} = \dfrac{\Cf \pi \alpha_s}{2} z_1 z_2 \hat{s}.
\end{align}
where $\Cf = (N_c^2-1)/(2 N_c)$.
The only comment which might be worth making on this simple calculation is that, as we have done in Section~\ref{sec:g*g*g_FR}, the average over the polarization states of the off-shell gluon has to be performed according to Eq.~\eqref{eq:PolarizationSum}.

Then, supplying the squared matrix element Eq.~\eqref{eq:M2qg*q} with the phase space volume element and the flux factor, given by Eqs.~\eqref{eq:dPhi1} and~\eqref{eq:phi} respectively, leads to the cross section of the off-shell process $q g^* \to q$ 
\begin{align}
\label{eq:sigmaqg*q}
\sigma \big|_{q g^* \rightarrow q} = \dfrac{\Cf}{\Ca} \dfrac{\sigma_0}{|\vec{p}_T|^2} \ \delta(1-\tau),
\end{align}
where $\sigma_0$ is given by Eq.~\eqref{eq:sigma0}. Note that this result is related to the cross section of $g^* g^* \to g$ Eq.~\eqref{eq:sigmag*g*g} by a simple change of the colour charge prefactor.

Finally, equipped with the results given by Eqs.~\eqref{eq:sigmag*g*g} and~\eqref{eq:sigmaqg*q}, we are now able to perform high-energy resummation of all partonic channels, to which topic the next section shall be devoted.

%%%%%%%%%%%%%%%%%%%%%%%%%%%%%%%%%%%%%%%%%%%%%%%%%%%%%%%%%%%%%%%%%%%%%%%
%%%%%%%%%%%%%%%%%%%%%%%%%%%%%%%%%%%%%%%%%%%%%%%%%%%%%%%%%%%%%%%%%%%%%%%
%%%%%%%%%%%%%%%%%%%%%%%%%%%%%%%%%%%%%%%%%%%%%%%%%%%%%%%%%%%%%%%%%%%%%%%

\section{Impact Factors and Resummed Results}
\label{sec:Results}

Thanks to the relations given by Eqs.~\eqref{eq:Relationqg} and~\eqref{eq:Relationqq}, the off-shell cross sections calculated in the previous section allow the evaluation of the high energy resummed transverse momentum distribution for all the following partonic processes:
\begin{subequations}
\label{eq:abc}
\begin{align}
g\(p_1\) + g\(p_2\) \to g\(p\) + X,\\
g\(p_1\) + q\(p_2\) \to g\(p\) + X,\\
q\(p_1\) + q\(p_2\) \to g\(p\) + X,\\
g\(p_1\) + q\(p_2\) \to q\(p\) + X,\\
q\(p_1\) + q\(p_2\) \to q\(p\) + X.
\end{align} 
\end{subequations}
To be specific, we shall calculate the dimensionless partonic differential distribution with respect to the transverse component $\pt$ of the momentum $p$, usually denoted by $\pt^4 d \hat{\sigma}/d\pt^2$.

Such a computation, as explained in Section~\ref{sec:HighEnergyResummation}, goes through the evaluation of particular impact factors, which in turn yield the resummed transverse momentum distributions via Eq.~\eqref{eq:relres}. However, when dealing with coloured final states, we need to take into account an extra subtlety: if the final state of the hard part contains gluons, in fact, we have to make sure that such gluons are indistinguishable from those contained in the ladders. 

We recall from Section~\ref{sec:HighEnergyResummation} that the derivation of high energy factorization of Refs.~\cite{Caola:2010,Muselli:2015} relies on the decomposition of the relevant observable of some final state $\mathcal{S}$ into hard and ladder parts. In so doing, the phase space of all ladder-radiated gluons is entirely contained in the ladder parts, whereas the hard part includes the phase space of $\mathcal{S}$. The indistinguishability of the ladder-radiated gluons is thus already accounted for: with $n-1$ emissions, for instance, is associated a factor of $1/(n-1)!$ in the ladder parts.

If the final state $\mathcal{S}$ contains one or more gluons as well, however, such decomposition eventually makes them distinguishable from those radiated off the ladders. For example, in the case where all the outgoing particles are gluons, each $\mathcal{O}\left(\alpha_s^n\right)$ term of the resummed observable should contain an overall combinatory factor of $1/n!$. As the hard part and ladders decomposition only accounts for a factor of $1/(n-1)!$, each $\mathcal{O}(\alpha_s^n)$ term has to be multiplied by an extra factor of $1/n$, so that the right gluon statistics is restored. In general, some complications arise if quarks are also present in the ladders, or if the hard part is not starting at $\Ord\(\as\)$, and hence contains particles which are different from gluons. 

We have proven a general formula, which allows the correction of the resummed result at all orders in $\as$ by acting on the impact factor $h$, evaluated keeping the particles produced in the hard part distinguishable from those radiated off the ladders, as
\begin{align}
\label{eq:GeneralCorrectionFormula}
\tilde{h}_{m,s}(N,M,M,\alpha_s) = - \int_0^M dM' \left[\dfrac{\chi_0(M)}{\chi_0(M')}\right]^{m-s} \dfrac{d \ln \chi_0(M')}{d M'}  h(N,M',M', \alpha_s),
\end{align}
where $\chi_0(M)$ is the LO BFKL kernel Eq.~\eqref{eq:chi0}. This correction is controlled by two natural numbers $m$ and $s$, which vary owing to the particular process under investigation. In particular, $m$ is the order in $\alpha_s$ of the hard part, whereas $s$ counts the number of quarks in the ladder emissions. Moreover, note that the simplifying choice $M_1=M_2=M$ in Eq.~\eqref{eq:GeneralCorrectionFormula} is motivated by the fact that, in the applications of interest to us, both $M$s ultimately have to be identified with the LL$x$ DGLAP anomalous dimension $\gamma\(N,\as\)$. For more details, as well as a complete derivation of Eq.~\eqref{eq:GeneralCorrectionFormula}, see Appendix~\ref{sec:ImpactFactorCorrection}.

With this tool in hand, we are now ready to compute the impact factors for all the processes of Eqs.~\eqref{eq:abc}. We shall first use the general definition Eq.~\eqref{eq:hpTdef} to evaluate $h$, and then correct it according to Eq.~\eqref{eq:GeneralCorrectionFormula} on a case by case basis. Particularly compact expressions for various $h$s could be obtained by selecting as hard scale $Q^2$ precisely the transverse modulus $\pt^2$ of the momentum $p^{\mu}$ of the tagged parton in Eqs.~\eqref{eq:abc}. In this case, the uncorrected impact factor $h$ can be defined, using $\sigma$ definitions Eq.~\eqref{eq:sigmag*g*g} and Eq.~\eqref{eq:sigmaqg*q}, and setting $\muf^2=\pt^2$, as
\begin{multline}
\label{eq:imp1}
\left. h\(N,M_1,M_2,\as\)\right|_{g g \to g}=M_1 M_2 R\(M_1\)R\(M_2\)\\
\int_0^\infty d\xi_1\, \xi_1^{M_1-1} \int_0^\infty d\xi_2\, \xi_2^{M_2-1}\int_0^{2\pi}\frac{d\theta}{2\pi} \pt^2\left.\sigma\right|_{g^* + g^* \to g} \\
\delta\(1-\xi_1-\xi_2-2\sqrt{\xi_1}\sqrt{\xi_2}\cos\theta\)
\end{multline}
in the $g + g \to g + X$ case, and 
\beq
\label{eq:imp2}
\left. h\(N,M,0,\as\)\right|_{g q \to q}=M R\(M\)\int_0^\infty d\xi\, \xi^{M-1}\pt^2\left.\sigma\right|_{g^* + q \to q} \delta\(1-\xi_1\)
\eeq
in the $g + q \to q + X$ case respectively. We denote by $\theta$ in Eq.~\eqref{eq:imp1} the angle between the directions of the transverse momenta $k_{1,\sss T}$ and $k_{2,\sss T}$ (see Eq.~\eqref{eq:Kinematics}). The impact factors for the other channels in Eqs.~\eqref{eq:abc} are recovered from these two by exploiting Eqs.~\eqref{eq:Relationqg} and~\eqref{eq:Relationqq}. 

Before we go any further, however, a clarification about this hard scale choice is due. At LL$x$ accuracy, the particular value of the hard scale is in fact meaningless, since any variation would only produce subleading effects. All the expressions we are going to present are thus valid all the same for any other choice of the hard scale. Note however that, even if at LL$x$ accuracy the consequences of such a choice are formally subleading, this does not mean that they should be negligible from a phenomenological point of view as well. Indeed, many phenomenological studies on fixed order expansions~\cite{Currie:2016bfm,Currie:2017ctp} show that jet production is particularly sensitive to the choice of this central scale. Nonetheless, since such effects are subleading in the high energy regime, our choice is only motivated by the relative simplicity of the presentation.

Then, by inserting Eqs.~\eqref{eq:sigmag*g*g} and~\eqref{eq:sigmaqg*q} for the off-shell cross sections into Eqs.~\eqref{eq:imp1} and~\eqref{eq:imp2}, performing the integrations, and finally applying the relations Eqs.~\eqref{eq:Relationqg} and~\eqref{eq:Relationqq}, we obtain the uncorrected impact factors for all the processes listed in Eqs.~\eqref{eq:abc}:
\begin{subequations}
\label{eq:impacts}
\begin{align}
\left.h\(N,M,M,\as\)\right|_{g g \to g} &= \sigma_0 \, M^2 \left[R(M)\right]^2 \dfrac{\left[\Gamma(M)\right]^2 \Gamma(1-2M)}{\left[\Gamma(1-M)\right]^2\Gamma(2M)},\\
\label{eq:hg*g*g}
\left.h\(N,M,M,\as\)\right|_{g q \to g} &= \sigma_0 \, M R(M)\left[M R(M)\dfrac{\left[\Gamma(M)\right]^2 \Gamma(1-2M)}{\left[\Gamma(1-M)\right]^2\Gamma(2M)}-1\right],\\
\left.h\(N,M,M,\as\)\right|_{q q \to g} &=\sigma_0 \, M R(M)\left[M R(M)\dfrac{\left[\Gamma(M)\right]^2 \Gamma(1-2M)}{\left[\Gamma(1-M)\right]^2\Gamma(2M)}-2\right],\\
\left.h\(N,M,0,\as\)\right|_{q g \to q} &=\frac{\Cf}{\Ca} \sigma_0 \, M R\(M\),\\
\left.h\(N,M,0,\as\)\right|_{q q \to q} &=\(\frac{\Cf}{\Ca}\)^2 \sigma_0 \, M R\(M\),
\end{align}
\end{subequations}
where we have put $M_1=M_2=M$ for the sake of simplicity.

From these expressions for the impact factors we can finally derive the small-$x$ resummed transverse momentum distributions for the processes listed in Eqs.~\eqref{eq:abc}, namely
\begin{subequations}
\label{eq:sigmares}
\begin{align}
\left.\pt^4\frac{d\sigma}{d\pt^2}\(N,\as\)\right|_{g + g \to g + X} &=\left.\tilde{h}_{1,0}\(N,\gamma\(N,\as\),\gamma\(N,\as\),\as\)\right|_{g g \to g},\\
\left.\pt^4\frac{d\sigma}{d\pt^2}\(N,\as\)\right|_{g + q \to g + X} &=\left.\tilde{h}_{1,1}\(N,\gamma\(N,\as\),\gamma\(N,\as\),\as\)\right|_{g q \to g},\\
\left.\pt^4\frac{d\sigma}{d\pt^2}\(N,\as\)\right|_{q + q \to g + X} &=\left.\tilde{h}_{1,2}\(N,\gamma\(N,\as\),\gamma\(N,\as\),\as\)\right|_{g g \to g},\\
\label{eq:q1}
\left.\pt^4\frac{d\sigma}{d\pt^2}\(N,\as\)\right|_{g + q \to q + X} &=\left.h\(N,\gamma\(N,\as\),0,\as\)\right|_{g q \to q},\\
\label{eq:q2}
\left.\pt^4\frac{d\sigma}{d\pt^2}\(N,\as\)\right|_{q + q \to q + X} &=\left.h\(N,\gamma\(N,\as\),0,\as\)\right|_{q q \to q},
\end{align}
\end{subequations}
where $\tilde{h}_{m,s}$ is defined according to Eq.~\eqref{eq:GeneralCorrectionFormula}, and $\gamma\(N,\as\)$ is again the LL$x$ DGLAP anomalous dimension. Note that quark transverse momentum distributions Eqs.~\eqref{eq:q1} and~\eqref{eq:q2} have not to be corrected, since no gluon is present in the final state of the hard part. Moreover, we recall that $N$ is always the Mellin variable conjugated to 
\beq
\label{eq:xdef}
x=\frac{\pt^2}{\hat{s}}.
\eeq 

Equations.~\eqref{eq:impacts} and~\eqref{eq:sigmares} allow the high-energy resummation at LL$x$ accuracy of the transverse momentum distribution of a generic gluon or quark in the final state, and ultimately represent the main results of this paper. In the following subsection we are going to present the series expansion of Eqs.~\eqref{eq:sigmares} at the first orders in $\as$, and discuss the analytic cross-checks against the corresponding fixed order evaluations.

%%%%%%%%%%%%%%%%%%%%%%%%%%%%%%%%%%%%%%%%%%%%%%%%%%%%%%%%%%%%%%%%%%

\subsection{Fixed order expansion}
\label{sec:check}
Starting from Eqs.~\eqref{eq:sigmares}, we perform the expansion in powers of $\as$, and compute the Mellin inverse transformation term by term, so as to reach an analytic result up to $\Ord\(\as^4\)$ in momentum space. Recalling the definition of $x$, given by Eq.~\eqref{eq:xdef}, and exploiting the results
\begin{align}
R\(M\)&= 1+ \frac{8}{3} \zeta(3) M^3 + \Ord\(M^4\),\\
\gamma\(N,\as\)&=\frac{\Ca}{\pi} \frac{\as}{N}+ 2\zeta(3) \frac{\Ca^4}{\pi^4} \frac{\as^4}{N^4} +\Ord\(\as^5\),
\end{align}
we obtain the following series expansions for the different partonic channel contributions:
\begin{subequations}
\label{eq:expansions}
\begin{align}
\left.\pt^4\frac{d\sigma}{d\pt^2}\(x,\as\)\right|_{g + g \to g + X}&=\sigma_0\(\frac{\Ca}{\pi}\as+ \frac{34}{45}\frac{\Ca^4}{\pi^4}\zeta(3) \as^4 \ln^3\frac{1}{x}+\Ord\(\as^5\)\),\\
\label{eq:SigmagggCorrected}
\left.\pt^4\frac{d\sigma}{d\pt^2}\(x,\as\)\right|_{g + q \to g + X}&=\sigma_0\(\frac{\Cf}{\pi}\as+ \frac{3}{4}\frac{\Cf\Ca^3}{\pi^4}\zeta(3) \as^4 \ln^3\frac{1}{x}+\Ord\(\as^5\)\),\\
\left.\pt^4\frac{d\sigma}{d\pt^2}\(x,\as\)\right|_{q + q \to g + X}&=\sigma_0\(\frac{20}{27}\frac{\Cf^2\Ca^2}{\pi^4}\zeta(3) \as^4 \ln^3\frac{1}{x}+\Ord\(\as^5\)\),\\
\left.\pt^4\frac{d\sigma}{d\pt^2}\(x,\as\)\right|_{g + q \to q + X}&=\sigma_0\(\frac{\Cf}{\pi}\as+ \frac{7}{9}\frac{\Cf\Ca^3}{\pi^4}\zeta(3) \as^4 \ln^3\frac{1}{x}+\Ord\(\as^5\)\),\\
\left.\pt^4\frac{d\sigma}{d\pt^2}\(x,\as\)\right|_{q + q \to q + X}&=\frac{\Cf}{\Ca}\sigma_0\(\frac{\Cf}{\pi}\as+ \frac{7}{9}\frac{\Cf\Ca^3}{\pi^4}\zeta(3) \as^4 \ln^3\frac{1}{x}+\Ord\(\as^5\)\).
\end{align}
\end{subequations}

We have cross-checked our predictions at $\Ord\(\as^2\)$ and $\Ord\(\as^3\)$\footnote{Note that $\sigma_0$ contains an extra power of $\as$, see Eq.~\eqref{eq:sigma0}.} against the LL$x$ behaviour of LO and NLO fixed order calculations. In particular, we highlight that, at LO, the predictions coming from $g + q \to g + X$ and $g + q \to q + X$ are identical as expected, since at $\Ord\(\as^2\)$ the quark and the gluon in the final state are back-to-back, and thus share the same transverse momentum behaviour. Instead, in the quark-quark channel, $q + q \to g + X$ and $q + q \to q + X$ predict different behaviours already at LO. This is well understood, since there are two possible final states in this case: $q + q \to g + g$ and $q + q \to q + q$. We have checked that the former matches the LO high energy prediction coming from $q + q \to g + X$, whereas the latter matches the LO high energy prediction coming from $q + q \to q + X$.

The comparison at $\Ord\(\as^3\)$ is far less trivial than at $\Ord\(\as^2\)$, since we have to consider more particles in the final state. To this end, we follow a technique similar to the one presented in Ref.~\cite{Muselli:2017ikh} in the case of Higgs boson production. The basic idea is that we compute directly the leading contributions in the high energy limit by retaining only the relevant Feynman diagrams. We present in details the comparison at $\Ord\(\as^3\)$ for the $g + g \to g + X$ case in Appendix~\ref{sec:DirectCalculation}. We have applied the same technique to check also all the other channels at $\Ord\(\as^3\)$.

%%%%%%%%%%%%%%%%%%%%%%%%%%%%%%%%%%%%%%%%%%%%%%%%%%%%%%%%%%%%%%%%%%%%%%%

\section{Conclusions and Outlook}
\label{sec:Conclusions}

In this paper we have discussed the generalization of high energy resummation to processes with strong-interacting final states. As a first application, we have worked out the high energy resummed transverse momentum distribution of the outgoing parton in all partonic processes of the form $g(q)+g(q)\to g(q)+X$ at LL$x$ accuracy. The main results are thus the impact factors given by Eqs.~\eqref{eq:impacts}, and the resummed transverse momentum distributions given by Eqs.~\eqref{eq:sigmares}. In order to reach this result, several new techniques have been exploited, together with the general analysis for the treatment of coloured final states discussed in Section~\ref{sec:RestoringGaugeInvariance}. 

We have shown in Section~\ref{sec:ANewApproach} that high energy resummation at LL$x$ accuracy can be achieved by exponentiating the LO matrix element of the off-shell hard process even in the case where such process is trivial or --~what is more~-- forbidden in conditions of on-shellness. Besides giving an insight into the theory of high energy resummation, this concept leads to a substantial simplification of the calculations, and we expect it will have future impact on the evaluation of high energy contributions to processes with jets in the final state.

Another key aspect of our analysis was colour management. 
Not only we have shown that working with coloured final states in a systematic and efficient way is possible, but we have also provided an explicit consistency cross-check of two recent techniques (presented in Refs.~\cite{Hameren, Kotko}) and the conventional method. Moreover, we have derived a general formula, given by Eq.~\eqref{eq:GeneralCorrectionFormula}, by means of which the indistinguishability of the outgoing gluons contained in the hard part from those radiated off the ladders can be restored systematically at all orders in $\as$. 

However, it is important to stress that the results achieved in this paper do not allow the evaluation of the LL$x$ behaviour of all the possible jet observables. Indeed, full inclusivity over the ladder emissions is for now mandatory in the derivation of high energy factorization, whereas some control over their kinematics as well would in general be required in the definition of many jet observables.

The natural next step is phenomenology. To this end, since we have only worked at parton level so far, the future plan consists in a thorough analysis of the effects of the various jet clustering algorithms and fragmentation functions, essential in order to build hadronic observables. 

Nevertheless, we wish to point out that for some observables, such as one-jet inclusive cross section and leading jet transverse momentum distribution, jet cluster analysis is in principle possible at high energy even requiring inclusiveness on the ladders phase space. Indeed, a particular order in the ladder emissions already exists in both transverse momentum and rapidity distributions~\cite{Caola:2010}. Therefore, for certain observables, and for particular jet algorithms, a complete determination of the final state kinematics is possible even with an exclusive control on the hard part alone at partonic level. Further studies are however needed and considered as a primary outlook.

Nonetheless, our partonic results are straight away applicable in PDFs fits thanks to recent developments in matched LL$x$ code implementation~\cite{Bonvini:2016wki,Bonvini:2017ogt}, and we expect them to be used in the next small-$x$ resummed global fit.

%%%%%%%%%%%%%%%%%%%%%%%%%%%%%%%%%%%%%%%%%%%%%%%%%%%%%%%%%%%%%%%%%%%%%%%
%%%%%%%%%%%%%%%%%%%%%%%%%%%%%%%%%%%%%%%%%%%%%%%%%%%%%%%%%%%%%%%%%%%%%%%
%%%%%%%%%%%%%%%%%%%%%%%%%%%%%%%%%%%%%%%%%%%%%%%%%%%%%%%%%%%%%%%%%%%%%%%

\acknowledgments{We are particularly grateful to Stefano Forte for many useful discussions in the early stages of this project and for a critical reading of the manuscript. This project has received funding from the European Research Council (ERC) under the European Union's Horizon 2020 research and innovation programme (grant agreement No 725110).}

%%%%%%%%%%%%%%%%%%%%%%%%%%%%%%%%%%%%%%%%%%%%%%%%%%%%%%%%%%%%%%%%%%%%%%%
%%%%%%%%%%%%%%%%%%%%%%%%%%%%%%%%%%%%%%%%%%%%%%%%%%%%%%%%%%%%%%%%%%%%%%%
%%%%%%%%%%%%%%%%%%%%%%%%%%%%%%%%%%%%%%%%%%%%%%%%%%%%%%%%%%%%%%%%%%%%%%%

\newpage

\appendix

\section{Wilson Lines}
\label{sec:WilsonLines}

We provide here a practical guide to diagrammatic calculation of gauge-invariant scattering amplitudes with external off-shell gluon legs in the Wilson lines formalism. We recall that by gauge invariance we mean that the matrix element fulfils the Abelian Ward identities Eq.~\eqref{eq:AbelianWardIdentity}. For a thorough discussion we refer to the original paper Ref.~\cite{Kotko}.

As well known, off-shell scattering amplitudes fail to be gauge-invariant in the sense of Eq.~\eqref{eq:AbelianWardIdentity} when the final state is strong-interacting, if the calculation is performed by means of the standard QCD Feynman rules alone. In order for gauge invariance to be restored, additional contributions are thus to be added, which can only come from additional, non-QCD Feynman rules. In this formalism, the new set of rules is derived from the insertion in the QCD matrix element of a suitably regularised and normalised straight infinite Wilson line operator.

\begin{figure}[t!]
\centering
\includegraphics[scale=0.75]{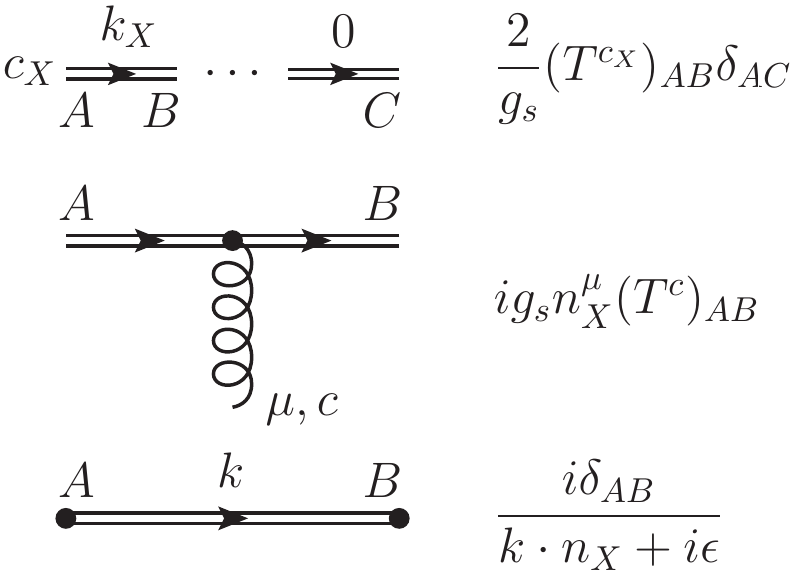}
\caption{Feynman rules for the insertion of a Wilson line operator corresponding to an incoming off-shell gluon with colour $c_X$, carrying momentum $k_X^{\mu} = z_X n_X^{\mu}+k^{\mu}_{X,\sss T}$. The top-most diagram is the so-called skeleton of the Wilson line; the dots denote potential gluon attachments according to the middle diagram, which represents the Wilson line-gluon coupling. The bottom diagram represents the Wilson line propagator.}
\label{fig:Wilson_FR}
\end{figure}

The calculation proceeds as follows. First, with each incoming off-shell gluon with momentum $k_X^{\mu}$ and colour $c_X$ is associated a so-called ``skeleton", namely the ends of the corresponding Wilson line, depicted on top of Figure~\ref{fig:Wilson_FR}. We recall that the incoming off-shell gluon momenta must have the usual high energy parametrization given by Eq.~\ref{eq:Kinematics}.

The skeleton is the most complicated object of this formalism: great care must be taken to ensure the right momentum and colour flow, so that the proper colour traces are recovered overall.
In the skeleton diagram depicted in Figure~\ref{fig:Wilson_FR}, for instance, momentum and colour flow to the right, so that the left-most side carries momentum $k_X^{\mu}$, whereas the right-most one has momentum zero. When more external off-shell gluon lines are involved, things get a little tricky. For our purposes, suffice it to say that, in the case of two off-shell incoming gluons, the direction of momentum and colour flow has to be reversed in one of the Wilson lines, as shown e.g. in Figure~\ref{fig:M2}. 
The general case of an arbitrary number of off-shell gluons is beyond the scope of this paper; see Ref.~\cite{Kotko} for the details.

Next, the dots in the skeleton diagram in Figure~\ref{fig:Wilson_FR} represent potential gluon attachments, which can be placed between the beginning and the end of the Wilson line according to the middle rule of Figure~\ref{fig:Wilson_FR}. To these gluon lines, then, the standard QCD Feynman rules have to be applied as required in order to produce the desidered final state. Also note that, as for the standard vertices, each Wilson line-gluon coupling comes with a momentum conservation constraint.

Finally, the proper Wilson line propagator has to be placed between subsequent gluon attachments, according to the lowest rule of Figure~\ref{fig:Wilson_FR}. The $+i \epsilon$ ($\epsilon>0$) term in the propagator denominator is used, as usual, to implement the Feynman boundary conditions.

One last remark on the analytic expressions of these Feynman rules is mandatory. Beware: the approach to the polarization vectors $e_X^{\mu}$ of the external off-shell gluons adopted in the Wilson lines formalism is different from the standard one presented in Section~\ref{sec:HighEnergyResummation}. Rather than treating them as proper polarization vectors, which have to be averaged over according to 
Eq.~\eqref{eq:PolarizationSum}, they here have the precise form
\begin{align}
\label{eq:EikonalCouplingAppendix}
e_X^{\mu} = z_X k_{X,\sss T} n_X^{\mu},
\end{align}
where $z_X$, $k_{X,\sss T}$ and $n_X^{\mu}$ are defined by the high energy parametrization of the off-shell gluon momentum $k_X^{\mu}$, given by Eq.~\eqref{eq:Kinematics}. The explicit expression of the off-shell polarization vectors given by Eq.~\eqref{eq:EikonalCouplingAppendix} is devised so as to automatically implement the average rule Eq.~\eqref{eq:PolarizationSum}, without the need of extra averaging factors. 

%%%%%%%%%%%%%%%%%%%%%%%%%%%%%%%%%%%%%%%%%%%%%%%%%%%%%%%%%%%%%%%%%%%%%%%
%%%%%%%%%%%%%%%%%%%%%%%%%%%%%%%%%%%%%%%%%%%%%%%%%%%%%%%%%%%%%%%%%%%%%%%
%%%%%%%%%%%%%%%%%%%%%%%%%%%%%%%%%%%%%%%%%%%%%%%%%%%%%%%%%%%%%%%%%%%%%%%

\section{High Energy Calculations: an Efficient Approach}
\label{sec:DirectCalculation}

The cross-check between the $\mathcal{O}(\alpha_s^3)$ term of the high energy resummed transverse momentum distribution of $g + g\rightarrow g + X$ Eq.~\eqref{eq:SigmagggCorrected} and the exact fixed-order calculation for $gg\rightarrow ggg$ is much more complicated than the $\mathcal{O}(\alpha_s^2)$ case, as it would also require the 1-loop corrections to $gg\rightarrow gg$ to be taken into account. The $\mathcal{O}(\alpha_s^3)$ squared scattering amplitudes for all $(2\rightarrow 2)$ and $(2\rightarrow 3)$ parton processes can be found in Ref.~\cite{EllisSexton}.

The squared matrix element, however, consists of hundreds of terms, which need to be integrated over a complicated phase space. Since an analytical cross-check would be preferable, a straightforward comparison with the exact result would be exceedingly cumbersome.

A possible solution comes from an alternative approach proposed in Refs.~\cite{Caola:2010,Muselli:2017ikh}: instead of taking the $x\rightarrow 0$ limit of the complete exact result, we are able to calculate directly the leading contributions in such limit, which we expect to be not only much simpler than the full expression, but indeed vanishing.

The basic insights at the heart of this technique are the same underlying the high energy factorization theorem: in an axial gauge, only $t$-channel gluon emissions contribute at LL$x$ accuracy~\cite{EllisGeorgiMachacek}, and interference and quark radiation are subleading effects~\cite{Muselli:2015}. The modus operandi basically consists in supplementing the fully differential cross section of the off-shell hard process with the small-$x$ enhanced phase space volume elements associated with as many ladder gluon emissions as desired. 

\begin{figure}[t!]
\centering
\includegraphics[scale=0.70]{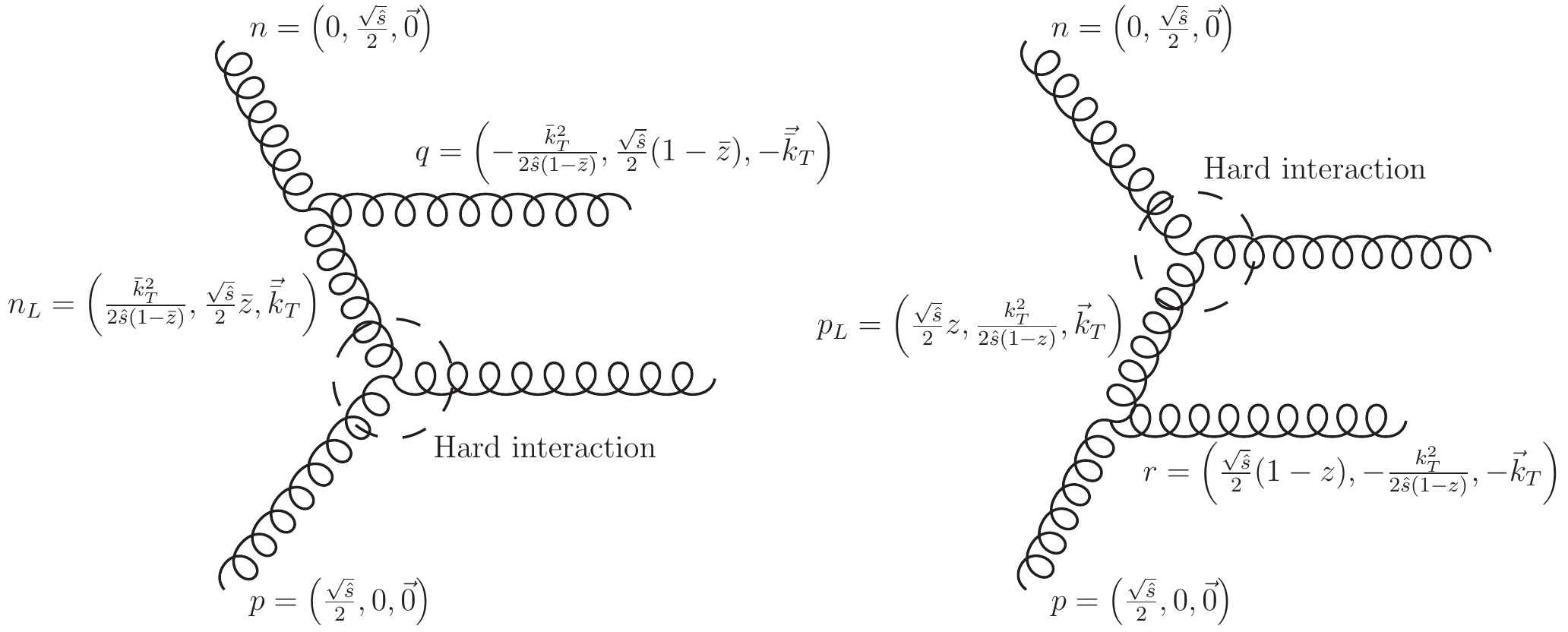}
\caption{Feynman diagrams giving the LL$x$ contributions to the single emission process in the small-$x$ limit. The relevant kinematics is displayed in terms of light-cone Sudakov components.}
\label{fig:SingleEmission}
\end{figure}

Let us first consider single gluon emission. The diagrams contributing at LL$x$ accuracy, together with the relevant kinematics, are depicted in Figure~\ref{fig:SingleEmission}. Up to power-suppressed terms in $z$ and $\kt^2/\hat{s}$, the LL$x$ contribution to the (dimensionless) transverse momentum distribution coming from single gluon radiation off the lower leg is
\begin{align}
\label{eq:sigma1down}
\pt^4\dfrac{d\tilde{\sigma}_{1-\text{down}}}{d \pt^2}(x; \epsilon) = \dfrac{1}{2!} \int \left[ \sigma_0 \delta\left( 1-\dfrac{x}{z}\right) \right] \left[ \bar{\alpha}_s \dfrac{d z}{z} \dfrac{d \xi}{\xi^{1+\epsilon}} \dfrac{(4 \pi)^{\epsilon}}{\Gamma(1-\epsilon)} \right] \delta(1-\xi) = \dfrac{1}{2} \sigma_0 \bar{\alpha}_s \left[ \dfrac{(4 \pi)^{\epsilon}}{\Gamma(1-\epsilon)} \right],	
\end{align}
where $\bar{\alpha}_s = \Ca \alpha_s/\pi$, $\xi$ is defined similarly to Eq.~\eqref{eq:xivariables}, and $\sigma_0$ is given by Eq.~\eqref{eq:sigma0}.

Some remarks about  Eq.~\eqref{eq:sigma1down} are in order. The tilde in Eq.~\eqref{eq:sigma1down} indicates that this observable might contain unsubtracted collinear singularities; the calculation is thus performed in $d = 4-2 \epsilon$ dimensions. In the first square bracket we recognize the dimensionless cross section of the hard interaction $gg\rightarrow g$ with only the lower incoming gluon left off-shell, which can be read from Eq.~\eqref{eq:sigmag*g*g}.
The second square bracket, instead, contains the small-$x$ enhanced phase space associated with gluon emission. The factors of $(4 \pi)^{\epsilon}/ \Gamma(1-\epsilon)$ coming from angular integrations are always subtracted in the $\overline{\text{MS}}$ subtraction scheme~\cite{Caola:2010}, and shall therefore be omitted hereafter. Lastly, the momentum conservation delta makes the observable a transverse momentum distribution, and the factor of $1/2!$ accounts for the two outgoing gluons being indistinguishable. Also in this case we define $x$ according to Eq.~\eqref{eq:xdef} by selecting as hard scale $Q^2=\pt^2$.

Due to simmetry, it is apparent that the contribution coming from the upper leg emission is the same as Eq.~\eqref{eq:sigma1down}. Thus, since interference is a subleading effect, the LL$x$ contribution in the small-$x$ limit to the unsubtracted (dimensionless) transverse momentum distribution of $gg\rightarrow gg$ simply reads
\begin{align}
\label{eq:sigma1unsub}
\pt^4\dfrac{d \tilde{\sigma}_{1} }{d \pt^2}(x; \epsilon) = \pt^4\dfrac{d \tilde{\sigma}_{1-\text{up}}}{d \pt^2}(x; \epsilon) + \pt^4\dfrac{d \tilde{\sigma}_{1-\text{low}}}{d \pt^2}(x; \epsilon) = \sigma_0 \bar{\alpha}_s.
\end{align}
It is trivial, in this case, to take the $\epsilon\rightarrow 0$ limit, and check that Eq.~\eqref{eq:sigma1unsub} matches both the $\mathcal{O}(\alpha_s^2)$ term of the resummed observable Eq.~\eqref{eq:SigmagggCorrected} and the small-$x$ limit of the full exact result, which can be found in any standard textbook.

\begin{figure}[t!]
\centering
\includegraphics[scale=0.6]{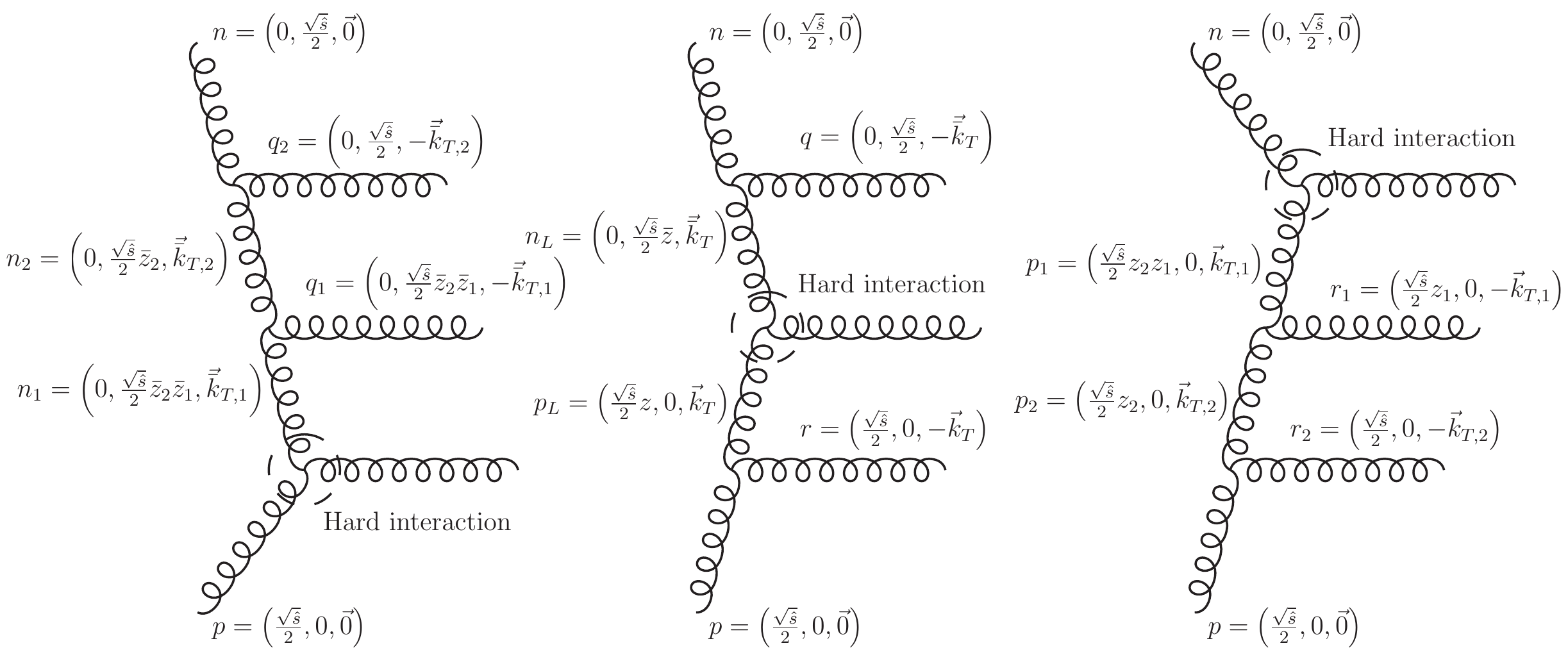}
\caption{Feynman diagrams giving the LL$x$ contributions to the double emission process in the small-$x$ limit. Crossed diagrams are omitted. The relevant kinematics is displayed in terms of light-cone Sudakov components in the high energy regime.}
\label{fig:DoubleEmission}
\end{figure}
Let us now tackle double emission. 
In Figure~\ref{fig:DoubleEmission} we can see the ladder-type diagrams contributing to this process at LL$x$ accuracy with the relevant kinematics. Note that the legs of the two emitted gluons also have to be exchanged, when they are both radiated off the same leg, thus leading to two crossed diagrams which we have not depicted in Figure~\ref{fig:DoubleEmission} for simplicity; it is however straightforward to prove that they give the same contributions as the corresponding uncrossed ones. We remember that each gluon emission is accompanied by a factor of $(4 \pi)^{\epsilon}/\Gamma(1-\epsilon)$, which is always subtracted in the $\overline{\text{MS}}$-scheme, and shall thus be omitted, in line with Refs.~\cite{Caola:2010,Muselli:2017ikh}.

Let us first consider double gluon emission from the same leg, say the lower one to make contact with the previous case. As small-$x$ leading contributions only come from the strongly ordered region $k_{1,\sss T}^2 \gg k_{2,\sss T}^2$~\cite{Catani:1990,Caola:2010}, the momentum conservation delta greatly simplifies
\begin{align}
\delta \left( 1- \xi_1 - \xi_2 - 2 \sqrt{\xi_1} \sqrt{\xi_2} \cos \theta \right) \approx \delta \left(1-\xi_1 \right),
\end{align}
where $\xi_i = k_{i,\sss T}^2/Q^2$, and $\theta$ is the angle between the directions of $\vec{k}_{1,\sss T}$ and $\vec{k}_{2, \sss T}$.
\\
Then, the generalization of Eq.~\eqref{eq:sigma1down} to double emission from the lower leg is
\begin{align}
\label{eq:sigma2down}
\pt^4\dfrac{d \tilde{\sigma}_{2-\text{down}}}{d \pt^2}(x; \epsilon) & = 2 \dfrac{1}{3!} \int \dfrac{d \theta}{2 \pi} \left[ \sigma_0 \delta\left(1-\dfrac{x}{z_1 z_2} \right) \right] \left[ \bar{\alpha}_s \dfrac{dz_2}{z_2} \dfrac{d \xi_2}{ \xi_2^{1+\epsilon}} \right] \left[ \bar{\alpha}_s \dfrac{dz_1}{z_1} \dfrac{d \xi_1}{ \xi_1^{1+\epsilon}}  \right] \delta(1-\xi_1) = \nonumber \\
& = \dfrac{1}{3} \sigma_0 \bar{\alpha}_s^2 \ln\left(\dfrac{1}{x}\right) \left[-\dfrac{1}{\epsilon}\right],
\end{align}
where the overall factor of $2$ takes account of the exchange of the emitted gluon legs as well.
Due to symmetry, double emission from the upper leg yields the same contribution as Eq.~\eqref{eq:sigma2down}.

Single emission from both legs is trickier: in line with Eq.~\eqref{eq:sigma2down}, its contribution at LL$x$ accuracy reads
\begin{align}
\pt^4\dfrac{d \tilde{\sigma}_{\text{up}-\text{down}}}{d \pt^2}(x; \epsilon) = \ & \dfrac{1}{3!} \int \dfrac{d \theta}{2 \pi} \left[\sigma_0 \delta \left( 1-\dfrac{x}{z \bar{z}}\right) \right] \left[ \bar{\alpha}_s \dfrac{dz}{z} \dfrac{d \xi}{\xi^{1+\epsilon}}\right] \left[ \bar{\alpha}_s \dfrac{d\bar{z}}{\bar{z}} \dfrac{d \bar{\xi}}{\bar{\xi}^{1+\epsilon}}\right] \times \nonumber \\
& \times \delta\left(1-\xi-\bar{\xi}-2 \sqrt{\xi} \sqrt{\bar{\xi}} \cos \theta \right),
\end{align}
where $\xi$ and $\bar{\xi}$ are defined as usual, but the angular dependence in the momentum conservation delta is not suppressed by strong ordering and has to be handled. Integrations can still be performed in closed form, yielding
\begin{align}
\label{eq:sigma2updown}
\pt^4\dfrac{d \tilde{\sigma}_{\text{up}-\text{down}}}{d \pt^2}(x; \epsilon) = \dfrac{1}{3!} \sigma_0 \bar{\alpha}_s^2 \ln \left(\frac{1}{x}\right) \dfrac{\Gamma(-\epsilon)^2 \Gamma(1+2 \epsilon)}{\Gamma(1+\epsilon)^2 \Gamma(-2 \epsilon)}  = \dfrac{1}{3} \sigma_0 \bar{\alpha}_s^2 \ln \left(\dfrac{1}{x} \right) \left[-\dfrac{1}{\epsilon} + \mathcal{O}\left(\epsilon^2\right) \right].
\end{align}
It is worth noting that the singular term of Eq.~\eqref{eq:sigma2updown} equals the contributions due to double emission from the same leg Eq.~\eqref{eq:sigma2down}.

Then, interference being subleading, the complete $\mathcal{O}(\alpha_s^3)$ unsubtracted (dimensionless) transverse momentum distribution of $gg\rightarrow ggg$ at LL$x$ accuracy is obtained by summing up the three contributions:
\begin{align}
\label{eq:sigma2unsubtracted}
\pt^4\dfrac{d \tilde{\sigma}_2}{d\pt^2}(x; \epsilon) = \sigma_0 \bar{\alpha}_s^2 \ln\left(\frac{1}{x}\right) \left[ - \dfrac{1}{\epsilon}+ \mathcal{O}(\epsilon^2)\right].
\end{align}

The last step consists in performing the $\overline{\text{MS}}$ subtraction of collinear singularities, which is most conveniently carried out in Mellin space in order to factorize convolutions. Indeed, as well known, the collinear radiation of a gluon produces a singular higher-order contribution which in $d = 4-2\epsilon$ dimensions has the form of a convolution of the lower order observable with the relevant splitting function, times a simple pole $-1/\epsilon$. Having already neglected the finite terms coming from angular integrations, collinear singularities are removed by subtracting a contribution of such form for each emitted gluon and then taking the $\epsilon \rightarrow 0$ limit
\begin{align}
\label{eq:MSSubtraction}
\pt^4\dfrac{d \sigma_{2}}{d \pt^2}(N) = \lim_{\epsilon \rightarrow 0 } \left\{ \dfrac{d \tilde{\sigma}_{2}}{d \xi_p}(N; \epsilon) - 2 \left[ \dfrac{1}{2!} \left(- \dfrac{1}{\epsilon} \right) P_{gg}^{(0)}(N) \dfrac{d \tilde{\sigma}_{1}}{d \xi_p}(N; \epsilon) \right] \right\},
\end{align}
where $\pt^4d \tilde{\sigma}_2/d\pt^2(N;\epsilon)$ and $\pt^4d \tilde{\sigma}_1/d\pt^2(N;\epsilon)$ are the Mellin transforms of Eqs.~\eqref{eq:sigma2unsubtracted} and~\eqref{eq:sigma1unsub} respectively, and the factor of $1/2!$ accounts for gluon indistinguishability. Finally, performing the Mellin transformations and substituting them into Eq.~\eqref{eq:MSSubtraction}, and recalling that at high energy~\cite{Jaroszewicz}
\begin{align}
P^{(0)}_{gg}(N) = \gamma^{(0)}_{gg}(N) = \dfrac{\bar{\alpha}_s}{N},
\end{align}
we see that the small-$x$ limit of the transverse momentum distribution of the outgoing gluon in the process $g + g\rightarrow g + X$ at $\Ord\(\as^3\)$ is indeed vanishing
\begin{align}
\pt^4\dfrac{d \sigma_{2}}{d \pt^2}(x) = \lim_{x \rightarrow 0} p_T^4 \dfrac{d \sigma}{d p_T^2}(x)\biggl|_{gg\rightarrow ggg} = 0,
\end{align}
in agreement with our prediction.

%%%%%%%%%%%%%%%%%%%%%%%%%%%%%%%%%%%%%%%%%%%%%%%%%%%%%%%%%%%%%%%%%%%%%%%
%%%%%%%%%%%%%%%%%%%%%%%%%%%%%%%%%%%%%%%%%%%%%%%%%%%%%%%%%%%%%%%%%%%%%%%
%%%%%%%%%%%%%%%%%%%%%%%%%%%%%%%%%%%%%%%%%%%%%%%%%%%%%%%%%%%%%%%%%%%%%%%

\section{Impact Factor Correction}
\label{sec:ImpactFactorCorrection}

In this appendix we present the complete derivation of Eq.~\eqref{eq:GeneralCorrectionFormula}.
Let us consider some resummed observable in Mellin-space of the form
\begin{align}
\label{eq:sigmaNa}
\sigma(N, \alpha_s) = \alpha_s^m \sum_{k=1}^{\infty} c_k \left(\dfrac{\alpha_s}{N}\right)^k,
\end{align}
which we want to modify so that each $\mathcal{O}\left(\alpha_s^{n}\right)$ term of the sum is divided by a factor of $n-s$ 
\begin{align}
\tilde{\sigma}(N, \alpha_s) = \alpha_s^m \sum_{k=1}^{\infty} \dfrac{c_k}{m+k-s} \left(\dfrac{\alpha_s}{N}\right)^k,
\end{align}
in such a way that one does not have to adjust each order separately. Since the sum in Eq.~\eqref{eq:sigmaNa} starts at $\mathcal{O}(\alpha_s^{m+1})$, the parameter $s$ is arbitrary apart from the constraint
\begin{align}
s < m+1.
\end{align}

For convenience, we define some reduced observables $\sigma(a)$ and $\tilde{\sigma}(a)$ by factorizing \\
$\sigma(N,\alpha_s)$ and $\tilde{\sigma}(N,\alpha_s)$ as
\begin{align}
\sigma(N,\alpha_s) & = \alpha_s^m \left(\dfrac{N}{\alpha_s}\right)^{m-s} \sum_{k=1}^{\infty} c_k \left(\dfrac{\alpha_s}{N}\right)^{k+m-s} = \dfrac{\alpha_s^m}{a^{m-s}} \sum_{k=1}^{\infty} c_k a^{k+m-s} \equiv \dfrac{\alpha_s^m}{a^{m-s}} \sigma(a), \nonumber \\ 
\tilde{\sigma}(N,\alpha_s) & = \frac{\alpha_s^m}{a^{m-s}} \sum_{k=1}^{\infty} \dfrac{c_k}{m+k-s} a^{m+k-s} \equiv \dfrac{\alpha_s^m}{a^{m-s}} \tilde{\sigma}(a), 
\end{align}
where we have introduced the variable
\begin{align}
a = \dfrac{\alpha_s}{N}
\end{align}
for the sake of simplicity. 

By using the definitions of $\sigma(a)$ and $\tilde{\sigma}(a)$, it is straightforward to prove that the latter must be a solution of the differential equation
\begin{align}
\label{eq:DiffEqSigma}
\dfrac{d \tilde{\sigma}}{d a}(a) = \dfrac{1}{a} \sigma(a), 
\end{align}
with initial condition
\begin{align}
\tilde{\sigma}(0) = 0.
\end{align}

Furthermore, let us introduce two new functions of $M$
\begin{align}
h(M) & := \sigma\left( a = \dfrac{1}{\chi_0(M)}\right), \\
\tilde{h}(M) & := \tilde{\sigma}\left( a = \dfrac{1}{\chi_0(M)}\right),
\end{align}
where $\chi_0(M)$ is the LO BFKL kernel given by Eq.~\eqref{eq:chi0}, which fulfils the BFKL duality relation~\cite{Altarelli:1999vw,Altarelli:2001ji}
\begin{align}
a \chi_0\left( M = \gamma_s(a) \right) = 1.
\end{align}

Indeed, these new functions are close relatives of the impact factors. Considering for instance $h(M)$, from thefollowing chain of equalities
\begin{align}
h\left( M \right)\biggl|_{M=\gamma_s(a)} & = \sigma \left(\dfrac{1}{\chi_0\left(M=\gamma_s(a)\right)} \right) = \sigma(a) = \dfrac{a^{m-s}}{\alpha_s^m} \sigma(N,\alpha_s) = \nonumber \\
& = \dfrac{1}{\alpha_s^m} \dfrac{1}{\left[\chi_0(M) \right]^{m-s}} h(0,M,M,\alpha_s) \biggl|_{M=\gamma_s(a)},
\end{align}
it follows that
\begin{align}
h(M) & = \dfrac{1}{\alpha_s^m} \dfrac{1}{\left[\chi_0(M) \right]^{m-s}} h(0,M,M,\alpha_s).
\end{align}
The same relation also holds for $\tilde{h}(M)$
\begin{align}
\tilde{h}(M) & = \dfrac{1}{\alpha_s^m} \dfrac{1}{\left[\chi_0(M) \right]^{m-s}} \tilde{h}(0,M,M,\alpha_s),
\end{align}
where we have introduced the corrected impact factor $\tilde{h}$, which will be our final desired result.

Changing variable from $a$ to $1/\chi_0(M)$ into Eq.~\eqref{eq:DiffEqSigma} and turning the derivative with respect to $1/\chi_0(M)$ into a derivative with respect to $M$ then leads to a differential equation for $\tilde{h}(M)$
\begin{align}
\label{eq:EqDiffhTilde}
\dfrac{d}{dM} \tilde{h}(M) = -\dfrac{1}{\chi_0(M)} \dfrac{d \chi_0(M)}{dM} h(M),
\end{align}
whose solution with some suitable initial condition in $M_0$ is
\begin{align}
\label{eq:GenericSolutionhTilde}
\tilde{h}(M) = \tilde{h}(M_0) - \int_{M_0}^{M} dM' \dfrac{1}{\chi_0(M')} \left(\dfrac{d \chi_0(M')}{dM'} \right) h(M').
\end{align}

As far as the $g^*g^*\rightarrow g$ case is concerned, $m=1$ and we know from the explicit calculation Eq.~\eqref{eq:hg*g*g} that
\begin{align}
h(M) = \dfrac{1}{\alpha_s} \dfrac{1}{\left[\chi_0(M)\right]^{1-s}} h_{\pt}(0,M,M,\alpha_s) \underset{M\rightarrow 0}{\sim} M^{1-s} \left(\frac{\pi}{N_c}\right)^{1-s} \left[2 \dfrac{\sigma_0}{\alpha_s} M + \mathcal{O}\left(M^4\right) \right].
\end{align}
As a result, we can infer the small-$M$ behaviour of $\tilde{h}(M)$ to be
\begin{align}
\tilde{h}(M) \underset{M\rightarrow 0}{\sim} M^{1-s} \left(\frac{\pi}{N_c}\right)^{1-s} \left[\dfrac{2}{2-s} \dfrac{\sigma_0}{\alpha_s} M + \mathcal{O}\left(M^4\right) \right].
\end{align}
Therefore, if $s<m+1=2$,
\begin{align}
\lim_{M\rightarrow 0} \tilde{h}(M) = 0.
\end{align}

It is thus convenient to choose $\tilde{h}(0)=0$ as the initial condition to solve Eq.~\eqref{eq:EqDiffhTilde}. However, before letting $M_0=0$ in Eq.~\eqref{eq:GenericSolutionhTilde}, we have to make sure that the relevant integration is well defined on the domain $[0,M]$. This is immediately done by calculating the small-$M$ behaviour of the integrand function
\begin{align}
-\dfrac{1}{\chi_0(M')} \dfrac{d \chi_0(M')}{dM'} h(M') \underset{M'\rightarrow 0}{\sim} M'^{1-s} \left(\dfrac{\pi}{N_c}\right)^{1-s} \left[ 2 \frac{\sigma_0}{\alpha_s} + \mathcal{O}(M') \right],
\end{align}
which is integrable on $[0,M]$, provided that $s<2$.

In conclusion, letting $M_0=0$ in Eq.~\eqref{eq:GenericSolutionhTilde} and writing $h(M)$ and $\tilde{h}(M)$ in terms of the corresponding impact factors gives the general correction formula
\begin{align}
\label{eq:GeneralCorrectionFormula0}
\tilde{h}(0,M,M,\alpha_s) = - \int_0^M dM' \left[\dfrac{\chi_0(M)}{\chi_0(M')}\right]^{m-s} \left(\dfrac{d \ln \chi_0(M')}{d M'} \right) h(0,M',M', \alpha_s).
\end{align}
Note that, although in this appendix we have set the explicit dependence on $N$ of the impact factors to $0$, since this is the only leading contribution at LL$x$ accuracy \cite{Muselli:2015}, the derivation works for general $N$ as well. We have thus proven the correction formula given in the text by Eq.~\eqref{eq:GeneralCorrectionFormula}.

%%%%%%%%%%%%%%%%%%%%%%%%%%%%%%%%%%%%%%%%%%%%%%%%%%%%%%%%%%%%%%%%%%%%%%%
%%%%%%%%%%%%%%%%%%%%%%%%%%%%%%%%%%%%%%%%%%%%%%%%%%%%%%%%%%%%%%%%%%%%%%%
%%%%%%%%%%%%%%%%%%%%%%%%%%%%%%%%%%%%%%%%%%%%%%%%%%%%%%%%%%%%%%%%%%%%%%%


\begin{thebibliography}{99}

\bibitem{Lipatov:1976}
  L.~N.~Lipatov,
  ``\textit{Reggeization of the Vector Meson and the Vacuum Singularity in Nonabelian Gauge Theories},''
  Sov.\ J.\ Nucl.\ Phys.\  {\bf 23} (1976) 338
   [Yad.\ Fiz.\  {\bf 23} (1976) 642].

\bibitem{Fadin:1975}
  V.~S.~Fadin, E.~A.~Kuraev and L.~N.~Lipatov,
  ``\textit{On the Pomeranchuk Singularity in Asymptotically Free Theories},''
  Phys.\ Lett.\  {\bf 60B} (1975) 50.
  doi:10.1016/0370-2693(75)90524-9.

\bibitem{Catani:1990}
  S.~Catani, M.~Ciafaloni and F.~Hautmann,
  ``\textit{Gluon Contributions to Small x Heavy Flavor Production},''
  Phys.\ Lett.\ B {\bf 242} (1990) 97.
  doi:10.1016/0370-2693(90)91601-7.

\bibitem{CataniHautmann}
  S.~Catani and F.~Hautmann,
  ``\textit{High-energy factorization and small x deep inelastic scattering beyond leading order},''
  Nucl.\ Phys.\ B {\bf 427} (1994) 475,
  doi:10.1016/0550-3213(94)90636-X
  [hep-ph/9405388].
  
    \bibitem{CataniCiafaloniHautmann}
  S.~Catani, M.~Ciafaloni and F.~Hautmann,
  ``\textit{High-energy factorization and small x heavy flavor production},''
  Nucl.\ Phys.\ B {\bf 366} (1991) 135,
  doi:10.1016/0550-3213(91)90055-3.
  

\bibitem{Caola:2010}
  F.~Caola, S.~Forte and S.~Marzani,
  ``\textit{Small x resummation of rapidity distributions: The Case of Higgs production},''
  Nucl.\ Phys.\ B {\bf 846} (2011) 167,
  doi:10.1016/j.nuclphysb.2011.01.001
  [arXiv:1010.2743 [hep-ph]].

\bibitem{Muselli:2015}
  S.~Forte and C.~Muselli,
  ``\textit{High energy resummation of transverse momentum distributions: Higgs in gluon fusion},''
  JHEP {\bf 1603} (2016) 122,
  doi:10.1007/JHEP03(2016)122
  [arXiv:1511.05561 [hep-ph]].

\bibitem{Hameren}
  A.~van Hameren,
  ``\textit{Helicity amplitudes for high-energy scattering processes},''
  PoS RADCOR {\bf 2013} (2013) 036.

\bibitem{HamerenKotkoKutak}
  A.~van Hameren, P.~Kotko and K.~Kutak,
  ``\textit{Helicity amplitudes for high-energy scattering},''
  JHEP {\bf 1301} (2013) 078,
  doi:10.1007/JHEP01(2013)078
  [arXiv:1211.0961 [hep-ph]].
  
\bibitem{Kotko}
  P.~Kotko,
  ``\textit{Wilson lines and gauge invariant off-shell amplitudes},''
  JHEP {\bf 1407} (2014) 128,
  doi:10.1007/JHEP07(2014)128
  [arXiv:1403.4824 [hep-ph]].

\bibitem{Lipatov1}
  L.~N.~Lipatov,
  ``\textit{Gauge invariant effective action for high-energy processes in QCD},''
  Nucl.\ Phys.\ B {\bf 452} (1995) 369,
  doi:10.1016/0550-3213(95)00390-E
  [hep-ph/9502308].
  
  %\cite{Moch:2004pa}
\bibitem{Moch:2004pa}
  S.~Moch, J.~A.~M.~Vermaseren and A.~Vogt,
  ``\textit{The Three loop splitting functions in QCD: The Nonsinglet case},''
  Nucl.\ Phys.\ B {\bf 688} (2004) 101
  doi:10.1016/j.nuclphysb.2004.03.030
  [hep-ph/0403192].
  %%CITATION = doi:10.1016/j.nuclphysb.2004.03.030;%%
  %866 citations counted in INSPIRE as of 20 Oct 2017
  
  %\cite{Vogt:2004mw}
\bibitem{Vogt:2004mw}
  A.~Vogt, S.~Moch and J.~A.~M.~Vermaseren,
  ``\textit{The Three-loop splitting functions in QCD: The Singlet case},''
  Nucl.\ Phys.\ B {\bf 691} (2004) 129
  doi:10.1016/j.nuclphysb.2004.04.024
  [hep-ph/0404111].
  %%CITATION = doi:10.1016/j.nuclphysb.2004.04.024;%%
  %679 citations counted in INSPIRE as of 20 Oct 2017
  
  \bibitem{EllisSexton}
  R.~K.~Ellis and J.~C.~Sexton,
  ``\textit{QCD Radiative Corrections to Parton Parton Scattering},''
  Nucl.\ Phys.\ B {\bf 269} (1986) 445,
  doi:10.1016/0550-3213(86)90232-4.
  
  %\cite{Nagy:2003tz}
\bibitem{Nagy:2003tz}
  Z.~Nagy,
  ``\textit{Next-to-leading order calculation of three jet observables in hadron hadron collision},''
  Phys.\ Rev.\ D {\bf 68} (2003) 094002
  doi:10.1103/PhysRevD.68.094002
  [hep-ph/0307268].
  %%CITATION = doi:10.1103/PhysRevD.68.094002;%%
  %504 citations counted in INSPIRE as of 20 Oct 2017
  
  %\cite{Currie:2016bfm}
\bibitem{Currie:2016bfm}
  J.~Currie, E.~W.~N.~Glover and J.~Pires,
  ``\textit{Next-to-Next-to Leading Order QCD Predictions for Single Jet Inclusive Production at the LHC},''
  Phys.\ Rev.\ Lett.\  {\bf 118} (2017) no.7,  072002
  doi:10.1103/PhysRevLett.118.072002
  [arXiv:1611.01460 [hep-ph]].
  %%CITATION = doi:10.1103/PhysRevLett.118.072002;%%
  %34 citations counted in INSPIRE as of 20 Oct 2017
  
    %\cite{Currie:2017ctp}
\bibitem{Currie:2017ctp}
  J.~Currie, E.~W.~N.~Glover, T.~Gehrmann, A.~Gehrmann-De Ridder, A.~Huss and J.~Pires,
  ``\textit{Single Jet Inclusive Production for the Individual Jet $p_{\rm T}$ Scale Choice at the LHC},''
  Acta Phys.\ Polon.\ B {\bf 48} (2017) 955
  doi:10.5506/APhysPolB.48.955
  [arXiv:1704.00923 [hep-ph]].
  %%CITATION = doi:10.5506/APhysPolB.48.955;%%
  %7 citations counted in INSPIRE as of 20 Oct 2017

\bibitem{Bonvini:2016}
  M.~Bonvini,
  ``\textit{Resummations in PDF fits},''
  PoS DIS {\bf 2016} (2016) 030
  [arXiv:1611.01925 [hep-ph]].

%\cite{Ball:2017otu}
\bibitem{Ball:2017otu} 
  R.~D.~Ball, V.~Bertone, M.~Bonvini, S.~Marzani, J.~Rojo and L.~Rottoli,
  ``\textit{Parton distributions with small-x resummation: evidence for BFKL dynamics in HERA data,}''
  arXiv:1710.05935 [hep-ph].
  %%CITATION = ARXIV:1710.05935;%%  
  
  \bibitem{Bonvini:2017ogt}
  M.~Bonvini, S.~Marzani and C.~Muselli,
  ``\textit{Towards parton distribution functions with small-$x$ resummation: HELL 2.0},''
  arXiv:1708.07510 [hep-ph].
  %%CITATION = ARXIV:1708.07510;%%
  
  \bibitem{Caola:2016}
  F.~Caola, S.~Forte, S.~Marzani, C.~Muselli and G.~Vita,
  ``\textit{The Higgs transverse momentum spectrum with finite quark masses beyond leading order},''
  JHEP {\bf 1608} (2016) 150,
  doi:10.1007/JHEP08(2016)150
  [arXiv:1606.04100 [hep-ph]].
  

  
  \bibitem{EllisGeorgiMachacek}
  R.~K.~Ellis, H.~Georgi, M.~Machacek, H.~D.~Politzer and G.~G.~Ross,
  ``\textit{Perturbation Theory and the Parton Model in QCD},''
  Nucl.\ Phys.\ B {\bf 152} (1979) 285,
  doi:10.1016/0550-3213(79)90105-6.
  
  \bibitem{CurciFurmanskiPetronzio}
  G.~Curci, W.~Furmanski and R.~Petronzio,
  ``\textit{Evolution of Parton Densities Beyond Leading Order: The Nonsinglet Case},''
  Nucl.\ Phys.\ B {\bf 175} (1980) 27,
  doi:10.1016/0550-3213(80)90003-6.
  
  \bibitem{AltarelliBallForte}
  G.~Altarelli, R.~D.~Ball and S.~Forte,
  ``\textit{Small $x$ resummation and HERA structure function data},''
  Nucl.\ Phys.\ B {\bf 599} (2001) 383,
  doi:10.1016/S0550-3213(01)00023-2
  [hep-ph/0011270].
  
  %\cite{Altarelli:1999vw}
\bibitem{Altarelli:1999vw}
  G.~Altarelli, R.~D.~Ball and S.~Forte,
  ``\textit{Resummation of singlet parton evolution at small x},''
  Nucl.\ Phys.\ B {\bf 575} (2000) 313
  doi:10.1016/S0550-3213(00)00032-8
  [hep-ph/9911273].
  %%CITATION = doi:10.1016/S0550-3213(00)00032-8;%%
  %120 citations counted in INSPIRE as of 15 Aug 2017

  %\cite{Altarelli:2001ji}
\bibitem{Altarelli:2001ji}
  G.~Altarelli, R.~D.~Ball and S.~Forte,
  ``\textit{Factorization and resummation of small x scaling violations with running coupling},''
  Nucl.\ Phys.\ B {\bf 621} (2002) 359
  doi:10.1016/S0550-3213(01)00563-6
  [hep-ph/0109178].
  %%CITATION = doi:10.1016/S0550-3213(01)00563-6;%%
  %100 citations counted in INSPIRE as of 15 Aug 2017
  
  %\cite{Bonvini:2016wki}
\bibitem{Bonvini:2016wki}
  M.~Bonvini, S.~Marzani and T.~Peraro,
  ``\textit{Small-$x$ resummation from HELL},''
  Eur.\ Phys.\ J.\ C {\bf 76} (2016) no.11,  597
  doi:10.1140/epjc/s10052-016-4445-6
  [arXiv:1607.02153 [hep-ph]].
  %%CITATION = doi:10.1140/epjc/s10052-016-4445-6;%%
  %4 citations counted in INSPIRE as of 15 Aug 2017
  
  \bibitem{Marzani:2008}
  S.~Marzani and R.~D.~Ball,
  ``\textit{High Energy Resummation of Drell-Yan Processes},''
  Nucl.\ Phys.\ B {\bf 814} (2009) 246
  doi:10.1016/j.nuclphysb.2009.01.029
  [arXiv:0812.3602 [hep-ph]].
  %%CITATION = doi:10.1016/j.nuclphysb.2009.01.029;%%
  %47 citations counted in INSPIRE as of 15 Aug 2017
  
\bibitem{PeskinColour}
  M.~E.~Peskin,
  ``\textit{Simplifying Multi-Jet QCD Computation},''
  arXiv:1101.2414 [hep-ph].
 
\bibitem{Dixon}
  L.~J.~Dixon,
  ``\textit{Calculating scattering amplitudes efficiently},''
  In *Boulder 1995, QCD and beyond* 539-582
  [hep-ph/9601359].

  %\cite{Muselli:2017ikh}
\bibitem{Muselli:2017ikh}
  C.~Muselli,
  ``\textit{High-energy Resummation of Higgs $p_{\mathrm {T}}$ Distribution},''
  Acta Phys.\ Polon.\ B {\bf 48} (2017) 1105.
  doi:10.5506/APhysPolB.48.1105
  %%CITATION = doi:10.5506/APhysPolB.48.1105;%%

\bibitem{Jaroszewicz}
  T.~Jaroszewicz,
  \textit{``Gluonic Regge Singularities and Anomalous Dimensions in QCD},''
  Phys.\ Lett.\  {\bf 116B} (1982) 291,
  doi:10.1016/0370-2693(82)90345-8.
    
  
  
  



\end{thebibliography}
\end{document}